\documentclass[lettersize,journal]{IEEEtran}
\IEEEoverridecommandlockouts

\usepackage{cite}
\usepackage[table,xcdraw]{xcolor}
\usepackage{graphicx,subfig,xcolor,setspace,amsthm,amsmath,amsfonts,cite,booktabs,epstopdf, gensymb, dblfloatfix,blindtext,amsthm, booktabs, arydshln,cuted}
\usepackage{graphicx}
\usepackage{textcomp}
\usepackage{xcolor}
\def\BibTeX{{\rm B\kern-.05em{\sc i\kern-.025em b}\kern-.08em
		T\kern-.1667em\lower.7ex\hbox{E}\kern-.125emX}}
\usepackage{subfig}
\usepackage{ upgreek }
\usepackage{graphicx}
\usepackage{graphics}
\usepackage{epstopdf}
\usepackage{amsmath}
\usepackage{multirow}
\usepackage{bm}
\usepackage{enumerate}
\usepackage{mathdots}
\usepackage{todonotes}
\usepackage{algorithmicx}
\usepackage{algpseudocode}
\usepackage{algorithm}
\usepackage{verbatim}
\usepackage{balance}
\usepackage{cite}
\usepackage[hyphens]{url}
\usepackage{acro}
\usepackage{color}
\usepackage{adjustbox}
\usepackage{esvect}
\usepackage{upgreek}
\usepackage{subfig}

\usepackage{microtype}

\def\Ntx{N_{\rmt}^{\rm x}}
\def\Nty{N_{\rmt}^{\rm y}}
\def\Nrx{N_{\rmr}^{\rm x}}
\def\Nry{N_{\rmr}^{\rm y}}
\newcommand{\sprl}{\shortparallel}
\definecolor{lightgray}{HTML}{b8b8aa}
\definecolor{newRed}{HTML}{bc3908}
\definecolor{purple(x11)}{rgb}{0.63, 0.36, 0.94}
\definecolor{cadmiumgreen}{rgb}{0.0, 0.42, 0.24}

\newcommand{\deh}[1]{\hspace{#1 mm}}
\usepackage{amsmath,amssymb}

\newcommand{\bbC}{{\mathbb{C}}}

\newcommand{\bbE}{{\mathbb{E}}}

\newcommand{\bbN}{{\mathbb{N}}}

\newcommand{\bbR}{{\mathbb{R}}}

\newcommand{\ba}{{\mathbf{a}}}
\newcommand{\bb}{{\mathbf{b}}}

\newcommand{\bi}{{\mathbf{i}}}
\newcommand{\bj}{{\mathbf{j}}}

\newcommand{\bn}{{\mathbf{n}}}
\newcommand{\bo}{{\mathbf{o}}}
\newcommand{\bp}{{\mathbf{p}}}

\newcommand{\br}{{\mathbf{r}}}
\newcommand{\bs}{{\mathbf{s}}}
\newcommand{\bt}{{\mathbf{t}}}

\newcommand{\bx}{{\mathbf{x}}}
\newcommand{\by}{{\mathbf{y}}}

\newcommand{\bA}{{\mathbf{A}}}
\newcommand{\bB}{{\mathbf{B}}}

\newcommand{\bF}{{\mathbf{F}}}

\newcommand{\bH}{{\mathbf{H}}}
\newcommand{\bI}{{\mathbf{I}}}

\newcommand{\bL}{{\mathbf{L}}}

\newcommand{\bN}{{\mathbf{N}}}

\newcommand{\bS}{{\mathbf{S}}}

\newcommand{\bW}{{\mathbf{W}}}
\newcommand{\bX}{{\mathbf{X}}}
\newcommand{\bY}{{\mathbf{Y}}}
\newcommand{\bZ}{{\mathbf{Z}}}
\newcommand{\rma}{{\mathrm{a}}}

\newcommand{\rmc}{{\mathrm{c}}}
\newcommand{\rmd}{{\mathrm{d}}}
\newcommand{\rme}{{\mathrm{e}}}

\newcommand{\rmh}{{\mathrm{h}}}

\newcommand{\rmo}{{\mathrm{o}}}
\newcommand{\rmp}{{\mathrm{p}}}

\newcommand{\rmr}{{\mathrm{r}}}
\newcommand{\rms}{{\mathrm{s}}}
\newcommand{\rmt}{{\mathrm{t}}}

\newcommand{\rmv}{{\mathrm{v}}}

\newcommand{\rmx}{{\mathrm{x}}}
\newcommand{\rmy}{{\mathrm{y}}}

\newcommand{\rmB}{{\mathrm{B}}}

\newcommand{\rmF}{{\mathrm{F}}}

\newcommand{\rmS}{{\mathrm{S}}}

\newcommand{\rmW}{{\mathrm{W}}}

\newcommand{\cN}{\mathcal{N}}
\newcommand{\cO}{\mathcal{O}}

\newcommand{\cS}{\mathcal{S}}

\newcommand{\sfi}{{\mathsf{i}}}

\newcommand{\sfk}{{\mathsf{k}}}

\newcommand{\bsfq}{\boldsymbol{\mathsf{q}}}
\newcommand{\bsfr}{\boldsymbol{\mathsf{r}}}

\newcommand{\sfT}{\mathsf{T}}

\newcommand{\bsfK}{\boldsymbol{\mathsf{K}}}

\newcommand{\bsfQ}{\boldsymbol{\mathsf{Q}}}
\newcommand{\bsfR}{\boldsymbol{\mathsf{R}}}

\newcommand{\bsfV}{\boldsymbol{\mathsf{V}}}
\newcommand{\bsfW}{\boldsymbol{\mathsf{W}}}

\newcommand{\bsfZ}{\boldsymbol{\mathsf{Z}}}
\newcommand{\balpha}{\boldsymbol{\alpha}}

\newcommand{\bgamma}{\boldsymbol{\gamma}}

\newcommand{\btheta}{\boldsymbol{\theta}}

\newcommand{\bvarpi}{\boldsymbol{\varpi}}

\newcommand{\bxi}{\boldsymbol{\xi}}

\newcommand{\bphi}{\boldsymbol{\phi}}

\newcommand{\bXi}{\boldsymbol{\Xi}}

\newcommand{\bUpsilon}{\boldsymbol{\Upsilon}}
\newcommand{\bPhi}{\boldsymbol{\Phi}}
\newcommand{\bPsi}{\boldsymbol{\Psi}}

\makeatletter
\def\munderbar#1{\underline{\sbox\tw@{$#1$}\dp\tw@\z@\box\tw@}}
\makeatother

\DeclareAcronym{3GPP}{
	short=3GPP,
	long=3rd Generation Partnership Project
}
\DeclareAcronym{ADC}{
	short=ADC,
	long=analog-to-digital converter
}
\DeclareAcronym{AMP}{
	short=AMP,
	long=approximate message passing
}
\DeclareAcronym{AoA}{
	short=AoA,
	long=angle-of-arrival
}
\DeclareAcronym{AoD}{
	short=AoD,
	long=angle-of-departure
}
\DeclareAcronym{APS}{
	short=APS,
	long=azimuth power spectrum
}
\DeclareAcronym{AWGN}{
	short=AWGN,
	long=additive white Gaussian noise
}
\DeclareAcronym{AV}{
	short=AV,
	long=autonomous vehicle
}
\DeclareAcronym{BS}{
	short=BS,
	long=base station
}
\DeclareAcronym{BP}{
	short=BP,
	long=belief propagation
}
\DeclareAcronym{BSM}{
	short=BSM,
	long=basic safety message
}
\DeclareAcronym{CDF}{
	short=CDF,
	long=cumulative distribution function
}
\DeclareAcronym{CIR}{
	short=CIR,
	long=channel impulse response
}
\DeclareAcronym{CNN}{
	short=CNN,
	long=convolutional neural network
}
\DeclareAcronym{ChanSTA}{
	short=ChanSTA,
	long=channel spatial and temporal attention
}

\DeclareAcronym{CP}{
	short=CP,
	long=cyclic-prefix
}

\DeclareAcronym{CS}{
	short=CS,
	long=compressed sensing
}
\DeclareAcronym{CSI}{
	short=CSI,
	long=channel state information
}

\DeclareAcronym{DFT}{
	short=DFT,
	long=discrete Fourier transform
}
\DeclareAcronym{DFS}{
	short=DFS,
	long=Doppler frequency shift
}
\DeclareAcronym{DL}{
	short=DL,
	long=deep learning
}

\DeclareAcronym{DNN}{
	short=DNN,
	long=deep neural network
}
\DeclareAcronym{DoA}{
	short=DoA,
	long=direction-of-arrival
}
\DeclareAcronym{DoD}{
	short=DoD,
	long=direction-of-departure
}
\DeclareAcronym{DSRC}{
	short=DSRC,
	long=dedicated short-range communication
}
\DeclareAcronym{EM}{
	short=EM,
	long=expectation maximization
}
\DeclareAcronym{EKF}{
	short=EKF,
	long=extended Kalman filter
}
\DeclareAcronym{FC}{
	short=FC,
	long=fully-connected
}

\DeclareAcronym{FDD}{
	short=FDD,
	long=frequency division duplex
}
\DeclareAcronym{FIM}{
	short=FIM,
	long=Fisher information matrix
}
\DeclareAcronym{FMCW}{
	short=FMCW,
	long=frequency modulated continuous wave
}

\DeclareAcronym{FoV}{
	short=FoV,
	long=field-of-view
}
\DeclareAcronym{GNSS}{
	short=GNSS,
	long=global navigation satellite system
}
\DeclareAcronym{GPS}{
	short=GPS,
	long=global positioning system
}
\DeclareAcronym{IoT}{
	short=IoT,
	long=internet of things
}
\DeclareAcronym{ISAC}{
	short=ISAC,
	long=integrated sensing and communication
}
\DeclareAcronym{IMU}{
	short=IMU,
	long=inertial measurement unit 
}
\DeclareAcronym{KL}{
	short=KL,
	long=Kullback–Leibler
}
\DeclareAcronym{KF}{
	short=KF,
	long=Kalman filter
}
\DeclareAcronym{LiDAR}{
	short=LiDAR,
	long=light detection and ranging
}
\DeclareAcronym{LOS}{
	short=LOS,
	long=line-of-sight
}
\DeclareAcronym{LPF}{
	short=LPF,
	long=low pass filter
}
\DeclareAcronym{LTE}{
	short=LTE,
	long=long term evolution
}
\DeclareAcronym{LS}{
	short=LS,
	long=least squares
}
\DeclareAcronym{LSTM}{
	short=LSTM,
	long=long short-term memory
}
\DeclareAcronym{mmWave}
{
	short = mmWave, 
	long = millimeter wave
}
\DeclareAcronym{MMSE}{
	short=MMSE,
	long=minimum mean squared error
}

\DeclareAcronym{MOMP}{
	short=MOMP,
	long=multidimensional orthogonal matching pursuit
}

\DeclareAcronym{MIMO}{
	short=MIMO,
	long=multiple-input multiple-output
}
\DeclareAcronym{MHA}{
	short=MHA,
	long=multi-head attention
}
\DeclareAcronym{MLE}{
	short=MLE,
	long=maximum likelihood estimation
}
\DeclareAcronym{MLP}{
	short=MLP,
	long=multilayer perceptron
}
\DeclareAcronym{MRR}{
	short=MRR,
	long=medium range radar
}
\DeclareAcronym{MPCs}{
	short=MPCs,
	long=multipath components
}
\DeclareAcronym{MSE}{
	short=MSE,
	long=mean squared error
}
\DeclareAcronym{NLOS}{
	short=NLOS,
	long=non-line-of-sight
}

\DeclareAcronym{NLP}{
	short=NLP,
	long=natural language processing
}

\DeclareAcronym{NR}{
	short=NR,
	long=new radio
}
\DeclareAcronym{OFDM}{
	short=OFDM,
	long=orthogonal frequency-division multiplexing
}
\DeclareAcronym{OMP}{
	short=OMP,
	long=orthogonal matching pursuit
}
\DeclareAcronym{PDP}{
	short=PDP,
	long=power delay profiles
}\DeclareAcronym{PDF}{
	short=PDF,
	long=probability distribution function
}
\DeclareAcronym{PDFs}{
	short=PDFs,
	long=probability distribution functions
}
\DeclareAcronym{PHD}{
	short=PHD,
	long=probability hypothesis density
}
\DeclareAcronym{PO}{
	short=PO,
	long=position and orientation
}
\DeclareAcronym{ppm}{
	short=ppm,
	long=parts-per-million
}
\DeclareAcronym{RF}{
	short=RF,
	long=radio frequency
 }
\DeclareAcronym{RMS}{
	short=RMS,
	long=root-mean-square
}
\DeclareAcronym{RPE}{
	short=RPE,
	long=relative precoding efficiency
}
\DeclareAcronym{RSU}{
	short=RSU,
	long=roadside unit
}
\DeclareAcronym{RTT}{
	short=RTT,
	long=round trip time
}
\DeclareAcronym{RX}{
	short=RX,
	long=receiver
}

\DeclareAcronym{RSRP}{
	short=RSRP,
	long=reference signal received power
 }
 \DeclareAcronym{RMSE}{
	short=RMSE,
	long=root mean squared error
 }

\DeclareAcronym{SNR}{
	short=SNR,
	long=signal-to-noise ratio
}
\DeclareAcronym{SLAM}{
	short=SLAM,
	long=simultaneous localization and mapping
}
\DeclareAcronym{SBL}{
	short=SBL,
	long=sparse Bayesian learning
}
\DeclareAcronym{SOMP}{
	short=SOMP,
	long=simultaneous orthogonal matching pursuit 
}
\DeclareAcronym{SOTA}{
	short=SOTA,
	long=state-of-the-art
}

\DeclareAcronym{TCN}{
	short=TCN,
	long=temporal convolutional network
}
\DeclareAcronym{TX}{
	short=TX,
	long=transmitter
}

\DeclareAcronym{TDoA}{
	short=TDoA,
	long=time-difference-of-arrival
}

\DeclareAcronym{ToA}{
	short=ToA,
	long=time-of-arrival
}

\DeclareAcronym{UL}{
	short=UL,
	long=uplink
}
\DeclareAcronym{ULA}{
	short=ULA,
	long=uniform linear array 
}
\DeclareAcronym{UKF}{
	short=UKF,
	long=unscented Kalman filter 
}
\DeclareAcronym{URA}{
	short=URA,
	long=uniform rectangular array 
}

\DeclareAcronym{V2I}{
	short=V2I,
	long=vehicle-to-infrastructure
}
\DeclareAcronym{V2V}{
	short=V2V,
	long=vehicle-to-vehicle
}
\DeclareAcronym{V2X}{
	short=V2X,
	long=vehicle-to-everything
}
\DeclareAcronym{VAs}{
	short=VAs,
	long=virtural anchors
}
\DeclareAcronym{VRU}{
	short=VRU,
	long=vulnerable road user
}
\DeclareAcronym{WLS}{
	short=WLS,
	long=weighted least squares
}
\DeclareAcronym{XR}{
	short=XR,
	long=extended reality
}

\definecolor{purple(x11)}{rgb}{0.63, 0.36, 0.94}
\definecolor{cadmiumgreen}{rgb}{0.0, 0.42, 0.24}

\usepackage{pifont}

\newcommand{\w}{{\mathbf{w}}_{(1)}}

\newcommand{\be}{\begin{eqnarray}}
\newcommand{\ee}{\end{eqnarray}}
\begin{document}
	\title{A Hybrid Model/Data-Driven Solution to Channel, Position and Orientation Tracking in mmWave Vehicular Systems}
	\author{Yun Chen,~\IEEEmembership{Student Member,~IEEE,} Nuria Gonz\'alez-Prelcic,~\IEEEmembership{Fellow,~IEEE,} Takayuki Shimizu,~\IEEEmembership{Member,~IEEE}, and Chinmay Mahabal,~\IEEEmembership{Member,~IEEE}
	\thanks{Y. Chen and N. Gonz\'alez-Prelcic are with the Department of Electrical and Computer Engineering, University of California, San Diego, CA 92161, USA (e-mail:\{yuc216, ngprelcic\}@ucsd.edu). T. Shimizu and C. Mahabal are with Toyota Motor North America, Mountain View, CA 94043, USA (e-mail: \{takayuki.shimizu, chinmay.mahabal\}@toyota.com).}}
	\maketitle
	
	\begin{abstract}
		Channel tracking in \ac{mmWave} vehicular systems is crucial for maintaining robust \ac{V2I} communication links, which can be leveraged to achieve high accuracy vehicle position and orientation tracking as a byproduct of communication. While prior work tends to simplify the system model by omitting critical system factors such as clock offsets, filtering effects, antenna array orientation offsets, and channel estimation errors, we address the challenges of a practical \ac{mmWave} \ac{MIMO} communication system between a single \ac{BS} and a vehicle while tracking the vehicle's \ac{PO} considering realistic driving behaviors. We first develop a channel tracking algorithm based on \ac{MOMP} with factoring (F-MOMP) to reduce computational complexity and enable high-resolution channel estimates during the tracking stage, suitable for \ac{PO} estimation. Then, we develop a network called VO-ChAT (Vehicle Orientation-Chanel Attention for orientation Tracking), which process the channel estimate sequence for orientation prediction. Afterward, a \ac{WLS} problem that exploits the channel geometry is formulated to create an initial estimate of the vehicle's 2D position. A second network named VP-ChAT (Vehicle Position-Channel Attention for position Tracking) refines the geometric position estimate.  VP-ChAT is a Transformer inspired network processing the historical channel and position estimates to provide the correction for the initial geometric position estimate. The proposed solution is evaluated using ray-tracing generated channels in an urban canyon environment. For 80\% of the cases it achieves a 2D position tracking accuracy of $26$ cm while orientation errors are kept below $0.5^\circ$.
	\end{abstract}
	
	\section{Introduction}
	
	The advancement of \ac{mmWave} \ac{MIMO} communication systems employing wide bandwidth and large antenna arrays enables high-resolution channel estimation, including accurate delay and angle acquisitions \cite{venugopal2017channel,palacios2022multidimensional}. Unlike lower frequency bands with dense \ac{MPCs} \cite{wang2024usrp, shakya2024urban}, mmWave channels exhibit sparsity and facilitate geometric localization  \cite{gonzalez2024integrated}. For example, in an outdoor vehicular scenario, the vehicle’s location can be derived from high-resolution estimates of the channel between the vehicle and a single \ac{BS} by exploiting geometric relationships between the path parameters and the locations of the scatterers, the \ac{BS}, and the vehicle \cite{gonzalez2024integrated}. Therefore, joint channel estimation and localization is a promising technology for real-world deployments as a cost-effective method for precise positioning while meeting the accuracy requirements of automated vehicles in various environments \cite{3GPPserviceReq}. While accurate single shot joint channel and position estimation for the {\em initial access} phase  (without including orientation) has been studied in our previous work \cite{chen2022joint, chen2023learning}, this paper focuses on reliable and high-accuracy vehicle \ac{PO} {\em tracking}. Sensor-based \ac{PO} tracking methods in vehicular settings utilizing \ac{IMU} \cite{or2024learning}, cameras \cite{ma2024vision, xu2025v2x}, \ac{LiDAR} \cite{lee2024lidar}, radar \cite{Abu2024Radar}, or sensor fusion \cite{wu2024aveh, Hashim2024uwb, li2024Multisensor}, are well-studied, but often suffer from compromised localization accuracy, e.g., with the \ac{GNSS} in urban canyons, or reduced reliability under adverse weather or lighting conditions. While solutions relying on mmWave communication signals are a promising alternative, \ac{SOTA} tracking solutions exploiting the link with a single \ac{BS} suffer from some limitations,  as discussed in Sec.~\ref{preWork}. These methods either fail to model the system realistically or do not achieve the desired localization accuracy for certain use cases when evaluated with practical \ac{mmWave}  communication channels and architectures.
	
	\subsection{Prior Work} \label{preWork}
	Representative channel-parameter-enabled position tracking methods are presented in \cite{Talvitie2023Orientation, Koivisto2022channel, gong2023high, zhao2023TDloc, gomez2023clock, Wang2024multipath, Bader2024leverag, Klus2024robust, Shamsesalehi2024abff,Venus2024graph, Leitinger2023data, gao2024message, Que2023joint, Yang2023angle, ge2022acomput, du2024general, Kaltiokallio2024integrated}. Two-stage approaches, in which channel parameters are acquired and subsequently employed for tracking, are studied in \cite{Talvitie2023Orientation, Koivisto2022channel, gong2023high, zhao2023TDloc, gomez2023clock, Wang2024multipath, Bader2024leverag, Klus2024robust}. In \cite{Talvitie2023Orientation},  antenna-level carrier phase measurements are used to acquire channel parameters including delays and angels that relate the \ac{PO} between the \ac{BS} and the \ac{XR} devices, and an \ac{EKF} is adopted to enable six-degrees-of-freedom (6DoF) tracking. However, the channel parameters are simulated using an error distribution function rather than estimated directly from received signals, which simplifies the complexity of real-world signal acquisition and processing. In contrast, channel parameter acquisition is included in \cite{Koivisto2022channel} and \cite{gong2023high}. In \cite{Koivisto2022channel}, optimal beam selection based on the \ac{FIM} is used to maximize the accuracy of delay and angle estimation, after which the channel parameters are tracked with an \ac{EKF} to obtain the user position, which is then tracked by another \ac{EKF}. Meanwhile, \cite{gong2023high} considers a high-speed outdoor vehicular scenario, addressing the complexity of estimating \ac{AoA}, \ac{ToA}, and Doppler shift using a sequential approach, and subsequently realizes localization by solving a \ac{WLS} optimization problem via Newton’s method. However, these methods require \ac{LOS} components and assume perfect synchronization between the \ac{TX} and \ac{RX} for the ranging purpose. Methods for tracking user \ac{PO} in both \ac{LOS} and \ac{NLOS} scenarios are proposed in \cite{zhao2023TDloc, gomez2023clock}. A tensor decomposition algorithm is proposed in \cite{zhao2023TDloc} to extract geometrical channel parameters and track the moving target by referencing the \ac{DFS} through intersecting virtual lines of estimated angles. The solution in \cite{gomez2023clock} introduces a \ac{CS}-based high-resolution multipath parameter estimation method, relating the known \ac{BS} \ac{PO} to the user \ac{PO} by solving a \ac{LS} estimation problem and nonlinear equations. However, both approaches neglect the filtering effects in the communication system and fail to address higher-order reflections, which negatively affect localization accuracy. Apart from the model-based solutions, approaches incorporating \ac{DL} are discussed in \cite{Wang2024multipath, Bader2024leverag, Klus2024robust, Shamsesalehi2024abff}. In \cite{Wang2024multipath}, a variational autoencoder architecture is proposed to extract position-related parameters such as \ac{TDoA} and \ac{AoA} from \ac{CIR} waveforms, mitigating errors in ranging and angles of \ac{NLOS} components, and the parameters are fused using a federated filter for user position tracking. 
	Without explicit channel parameter extraction approaches, \cite{Bader2024leverag} assumes the availability of ideally simulated channel parameters, and in \cite{Klus2024robust}, channel parameters are generated with uncertainties. In such cases, \cite{Bader2024leverag} introduces an ensemble-learning way to identify \ac{LOS} and single-bounce \ac{NLOS} components for geometrical localization, and adopt an \ac{UKF} together with supplemental odometer data to refine location estimates. A \ac{LSTM} \ac{DNN} is employed in \cite{Klus2024robust} to extract \ac{CSI} features in frequency and time domains and aggregates the information of \ac{ToA}, \ac{AoA}, and pair-wise received powers, for \ac{PO} tracking. Furthermore, a fingerprinting solution is presented in \cite{Shamsesalehi2024abff}, where the beamformed fingerprint data is input into a Transformer network to predict the user trajectories. 
	
	In addition to the two-stage strategies, joint channel and position tracking approaches are explored in \cite{Venus2024graph, Leitinger2023data, gao2024message, Que2023joint, Yang2023angle, ge2022acomput, du2024general, Kaltiokallio2024integrated}, leveraging the joint probability distribution of user \ac{PO} and channel multipath parameters, and employing various filtering methods for user state (PO) tracking. In \cite{Venus2024graph}, a factor graph is formulated with a sum-product algorithm (SPA) to calculate marginal posterior distributions of state variables including user \ac{PO} and channel parameters, enhancing \ac{NLOS} delay and amplitude estimates, followed by a particle-based implementation for state predictions. The factor graph with \ac{BP} methods are commonly applied to channel \ac{SLAM} \cite{Leitinger2023data, gao2024message}, modeling the state of users (e.g., \ac{PO}, velocity), physical anchors/BS, \ac{VAs}, and the geometry of \ac{MPCs}, with \ac{PDFs}. In \cite{Leitinger2023data}, where a super-resolution channel estimation algorithm is used to extract \ac{MPCs} and higher-order reflections are addressed by incorporating a ray-tracing module, a factor graph representation for the user state, anchor state, and channel measurements is established, a SPA is employed for belief calculation, and finally the user's PO are updated through a \ac{MMSE} estimator. In \cite{gao2024message}, where the factor graph construction remains similar, a continuous measurement correction method incorporating time-sequential measurements is integrated into the \ac{BP} process, enabling efficient message passing. Besides, in \cite{Que2023joint, Yang2023angle}, angle-based \ac{SLAM} are provided where the multipath angle estimates are acquired through beam sweeping. The user \ac{PO} state is obtained with \ac{IMU} inputs through particle filtering \cite{Que2023joint}, or a \ac{BP} framework to calculate  \ac{PDFs} of user states, the anchors, and channel measurements, followed by a \ac{MMSE} estimator to update user PO \cite{Yang2023angle}. While \cite{Que2023joint,Yang2023angle} rely on \ac{LOS} and first-order reflections, \cite{ge2022acomput, du2024general, Kaltiokallio2024integrated} address \ac{NLOS} situations, providing more comprehensive approaches considering multipath birth and disappearance. In \cite{ge2022acomput}, a Poisson multi-Bernoulli mixture density is used to represent the joint distribution of environmental landmarks including the anchors, and an \ac{EKF} is adopted to jointly update motion sensor and landmark states for user PO inference. Alternatively, as in \cite{du2024general}, landmark changes are predicted considering a birth \ac{PHD} added to the previous landmark \ac{PHD}, and particle filtering is adopted for updating user POs. In \cite{Kaltiokallio2024integrated}, a snapshot \ac{SLAM} method based on multipath geometry--excluding higher-order reflections--is incorporated to get the initial estimates for a multi-hypothesis linear filter, which contains a nearest neighbor filter for user state tracking and a \ac{PHD} filter for landmark tracking. 
	
	The above methods face several limitations. Many studies assume idealized channel multipath parameters without incorporating real-world channel estimation or tracking techniques \cite{Talvitie2023Orientation, Bader2024leverag, Shamsesalehi2024abff, gao2024message}. For approaches that include estimation or tracking of channel \ac{MPCs} \cite{Koivisto2022channel, gong2023high, zhao2023TDloc, gomez2023clock, Wang2024multipath, Klus2024robust, Leitinger2023data, Que2023joint,Yang2023angle, ge2022acomput, du2024general, Kaltiokallio2024integrated}, simplified communication systems are often considered by using \ac{ULA} instead of \ac{URA} at one or both ends to reduce processing complexity \cite{Koivisto2022channel, zhao2023TDloc, Klus2024robust, gomez2023clock, Leitinger2023data, Que2023joint, Yang2023angle, ge2022acomput, du2024general, Kaltiokallio2024integrated}, and filtering effects for time-domain channel processing are neglected \cite{gomez2023clock, gong2023high, Klus2024robust}. Furthermore, some methods do not address orientation tracking \cite{Koivisto2022channel, gong2023high, Wang2024multipath, Shamsesalehi2024abff}. \ac{PO} tracking algorithms relying on the presence of \ac{LOS} paths for ranging \cite{Talvitie2023Orientation, Koivisto2022channel, gong2023high, Leitinger2023data, Bader2024leverag} assume perfect synchronization between the \ac{TX} and \ac{RX} \cite{Koivisto2022channel, gong2023high, zhao2023TDloc, Venus2024graph, Leitinger2023data, du2024general}, or require \ac{RTT} measurements to cancel clock biases \cite{Wang2024multipath, Klus2024robust, Bader2024leverag}. In addition, algorithms that depend on LOS and/or first-order reflections \cite{zhao2023TDloc, Wang2024multipath, gao2024message, Que2023joint, Yang2023angle, ge2022acomput, du2024general} fail to address higher-order reflections negatively affecting the tracking performance. Most methods are evaluated in indoor environments  \cite{Talvitie2023Orientation, zhao2023TDloc, Venus2024graph, Leitinger2023data, gao2024message, Que2023joint, Yang2023angle, du2024general, Kaltiokallio2024integrated}, while the accuracy can degrade for outdoor complex scenarios. While \ac{DL}-based fingerprinting solutions achieve reasonable accuracy, e.g., \ac{RMSE} of $1\sim 2$ m \cite{Klus2024robust, Shamsesalehi2024abff}, they remain inadequate for achieving submeter-level tracking.

	\subsection{Contributions}    
	In this paper, we propose a novel hybrid model/data-driven framework for precise vehicle PO tracking as a byproduct of mmWave channel tracking. Considering a vehicle under tracking communicating with a roadside \ac{BS}, the approach starts with a low-complexity channel tracking algorithm, F-MOMP, to enable high-resolution channel estimates. To address the challenge of unknown vehicle orientations in realistic driving scenarios, which affect localization accuracy, we propose an attention-based network, VO-ChAT, to predict the current vehicle orientation with the input sequence of channel estimates. Following orientation compensation, the estimated channel paths are weighted for single-shot localization through a geometric transformation. Subsequently, we leverage historical channel and position information to enhance position estimation accuracy using a Transformer-inspired network, VP-ChAT, which shares a partial architecture with VO-ChAT in its encoder for processing the channel estimate sequence, while incorporating position estimates through its decoder to output the correction for the current single-shot position estimate. Our contributions are as follows:
	\begin{itemize}
		\item We consider a mmWave \ac{MIMO} communication system employing \ac{URA}s between a single \ac{BS} and a vehicle with the driver following realistic driving behavior models. The system model accounts for unknown clock offset drifts between the \ac{TX} and \ac{RX} and the system filtering effects. While conventional channel estimation algorithms fail due to the computational complexity associated with high-resolution channel estimation required for localization, we develop the F-MOMP algorithm (available at \cite{Chen_F-MOMP_2024}) for low-complexity and accurate channel tracking --with the delay accuracy of $0.1$ ns and angular accuracy of $2^\circ$ (at the 80th percentile)-- to enable vehicle localization through the estimated \ac{MPCs}.
		
		\item To address the unknown vehicle orientation incorporated in the estimated channel angular parameters that will affect the localization process, we design an attention-based network, VO-ChAT, to track the vehicle orientations. The network processes the input sequence of channel estimates to acquire the channel spatial and temporal evolution features, and concurrently integrates the historical orientation information to predict the current orientation. It achieves the orientation prediction error of $\leq 0.5^\circ$ for 80\% of the situations.
		
		\item After orientation compensation based on VO-ChAT predictions, we identify \ac{LOS} and first-order reflections referring to the vehicle's height obtained during the initial access phase, using which we implement the single-shot localization through channel path geometric transformations using a \ac{WLS} algorithm. Subsequently, we propose a Transformer-inspired network, VP-ChAT, to process channel and position estimate sequences. Specifically, a module structurally similar to VO-ChAT serves as the encoder for VP-ChAT to process the channel estimate sequence and extract channel spatial and temporal evolution features, while the decoder of VP-ChAT processes the current single-shot position estimate plus the position history as a query to determine the correction of the current position estimate. This approach achieves a position tracking accuracy of $0.15$ m at the 50th percentile and $0.43$ m at the 95th percentile.
		\item Our methods are evaluated using realistic ray-tracing simulated channels, generated based on snapshots captured along the vehicle's trajectory. The simulated channel database will be open sourced to provide the research community with a resource for evaluating new solutions to the joint channel and PO estimation/tracking problem using vehicular communication channels.        
	\end{itemize}
	
	The framework described in the paper is built upon the foundational design presented in our prior work \cite{chen2023sparse} with several key improvements. Vehicle trajectories are generated based on realistic driving behaviors, accounting for dynamic orientation changes which are tracked through the newly added VO-ChAT. The original channel tracking strategy based on MOMP is extended to the F-MOMP-based solution which incorporates the factoring operation to reduce computational complexity. The original single-shot localization through geometric transformations is refined as solving a \ac{WLS} estimation problem. The original V-ChAT network is tuned into VP-ChAT to accommodate updated vehicle trajectories for corrections of the single-shot position estimates. A larger and more comprehensive dataset is formed, and additional numerical experiments and comparisons with \ac{SOTA} studies are included.
	
	The rest of the paper is structured as follows: Sec. \ref{sec:sys_model} outlines the general \ac{V2I} communication setup, the driving behavior model, and the communication system model. Sec. \ref{sec:pos-track} details the stages of the proposed hybrid model/data-driven approach, including channel tracking, orientation prediction and compensation, single-shot localization, and position corrections leveraging historical channel and position estimates. Then, Sec. \ref{sec:SimResults} presents numerical results evaluating the proposed strategy and comparisons with prior work. Finally, Sec. \ref{conclu} concludes the paper by summarizing the key findings.
	
	\textbf{Notations:} $[\bx]_i$ and $[\bX]_{i,j}$ denote the $i$-th entry of a vector $\bx$ and the entry at $i$-th row and $j$-th column of a matrix $\bX$ (the same rule applies for a tensor). $\bX^{\sfT}$, $\bar{\bX}$, $\bX^*$, and $\bX^\dag$ are the transpose, conjugate, conjugate transpose, and pseudo inverse of $\bX$. $[\bX,\bY]$ and $[\bX;\bY]$ are the horizontal and vertical concatenation of $\bX$ and $\bY$. $\bX\otimes\bY$ and $\bX\odot\bY$ are the Kronecker product and Khatri-Rao product of $\bX$ and $\bY$. $\frak{X}\cup\frak{Y}$ is the union set of set $\frak{X}$ and $\frak{Y}$. $x\sim\cN(\dot{x},\sigma^2_x)$ denotes the variable $x$ follows the Gaussian distribution with mean $\dot{x}$ and variance $\sigma^2_x$.
	\begin{figure}
		\centering
		\includegraphics[width=\linewidth]{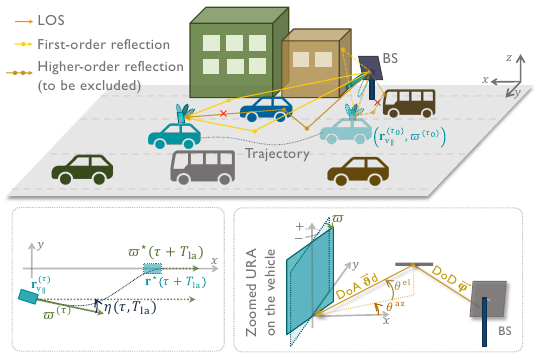}
		\caption{System model for tracking a vehicle in the urban canyon environment. }
		\label{sys_model}
	\end{figure}
	\begin{figure*}[!t]
		\centering
		\includegraphics[width=.9\linewidth]{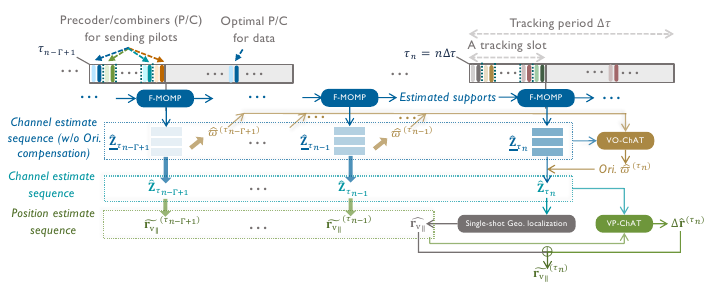}
		\caption{System diagram consisting of F-MOMP for channel tracking, VO-ChAT for vehicle orientation tracking, single-shot geometric (Geo.) localization using angles after orientation compensation, and VP-ChAT for vehicle position tracking.}
		\label{AlgPipeline}
	\end{figure*}
	\section{System Model}\label{sec:sys_model}
	We consider a mmWave vehicular communication system where an active car under tracking moves at the fast lane with the heading (orientation) changing according to driver behavior models \cite{van2023modeling}. The active vehicle starts from position $\br^{(\tau_0)}_{\rmv_\sprl}$, a two-dimensional vector comprising x and y coordinates, with the initial orientation $\varpi^{(\tau_0)}$.  At time $\tau$ when the vehicle is at position $\br_{\rmv_\sprl}^{(\tau)}$ with the orientation $\varpi^{(\tau)}$ and the moving speed $v_\rmv^{(\tau)}$, the driver checks the aiming point $\br^\star({\tau+T_{\rm la}})$ lying on the lane center line, where $T_{\rm la}$ is the looking ahead time depending on the environmental visibility, and steers the wheel continuously during the period from  $\tau$ to $\tau+T_{\rm la}$ expecting to arrive at $\br^\star({\tau+T_{\rm la}})$ with the orientation $\varpi^\star(\tau+T_{\rm la})$. Hence, the bearing angle from $\tau$ to $\tau+T_{\rm la}$, denoted as $\eta(\tau,T_{\rm la})$, is calculated by $\eta(\tau, T_{\rm la})=\varpi^\star(\tau+T_{\rm la})-\varpi^{(\tau)}$. We define the driver compensate control as $\delta(\tau)$, which needs to be increased meaning the driver controls the steering wheel with higher strength when $\eta(\tau,T_{\rm la})$ is large. Considering a continuous operation model, the following equations hold:
	\begin{align}
		&\eta(\tau, T_{\rm la}) = \int _{\tau}^{\tau+T_{\rm la}}\omega(t)\delta(t)\partial t;\label{bearAng}\\
		&\br^\star\deh{-.5}({\tau\deh{-1}+\deh{-1}T_{\rm la}})\deh{-1}-\br_{\rmv_\sprl}^{(\tau)}\deh{-1} = \deh{-1}\int_{\tau}^{\tau+T_{\rm la}}\deh{-1} [v_\rmv^{(\tau)}\cos(\varpi^{(t)}), v_\rmv^{(\tau)}\sin(\varpi^{(t)})]^\sfT\partial t,\label{distChange}
	\end{align}
	where $\omega(t)\sim\cN(\dot{\omega},\sigma^2_\omega)$ is the wheel steering rate at $t$, \eqref{bearAng} represents the cumulated orientation changes, and \eqref{distChange} represents the vehicle position changes. When implemented in the discrete domain, let $\Delta\tau=\tau_{n+1}-\tau_n$ be the sampling interval, vehicle's orientation and position can be updated as:
	\begin{align}
		&\delta(\tau_{n+1}) = K_\rme\left(\delta({\tau_{n}})+T_\rme \frac{\eta(\tau_{n+1},T_{\rm la}\deh{-1}-\deh{-1}\Delta\tau)-\eta(\tau_n,T_{\rm la})}{\Delta\tau}\right)\nonumber\\&\qquad\quad+n_\delta(\tau_{n+1});\\
		&\varpi^{(\tau_{n+1})} = \varpi^{(\tau_{n})}+\omega({\tau_n})\delta({\tau_n})\Delta\tau;\\
		&\br_{\rmv_\sprl}^{(\tau_{n+1})} = \br_{\rmv_\sprl}^{(\tau_n)} + \Delta\tau[v_\rmv^{(\tau_n)}\cos(\varpi^{(\tau_{n})}),v_\rmv^{(\tau_n)}\sin(\varpi^{(\tau_{n})})]^\sfT,
	\end{align}
	where $K_\rme$ and $T_\rme$ are the driver gain and leading time constants, $n_\delta(t)\sim \cN(0, \sigma^2_\delta)$ is the driver control noise being $\sigma^2_\delta$ the distribution variance.
	
	Downlink communication is performed between the active car and a single \ac{BS} at the roadside for tracking the channel and POs. The BS is equipped with a \ac{URA} of size $N_{\rmt}=\Ntx\times \Nty$ facing the road, and the vehicle has 4 smaller URAs placed vertically on the hardtop as in \cite{chen2022joint, chen2023learning}, each of which has a size of $N_{\rmr}=\Nrx\times \Nry$. A hybrid \ac{MIMO} communication architecture is adopted, with $N_\rmt^{\rm rf}$ and $N_\rmr^{\rm rf}$ \ac{RF} chains deployed at the \ac{TX} and \ac{RX}.  Hereby, the frequency selective mmWave channel containing $L$ \ac{MPCs} at a given time $\tau_n$ can be defined as 
	\begin{align}\label{chan_mod} \deh{-1}\bH_d^{(\tau_n)}\deh{-.8}=&\deh{-.5}\sum\limits_{\ell=1}^{L}\biggl(\deh{-.5}\alpha_\ell^{(\tau_n)} f_{\rmp}\left(dT_\rms-\left(t_\ell^{(\tau_n)}\deh{-.1}-\deh{-.8}t_{\rm off}^{(\tau_n)}\right)\right)\cdot\nonumber\\
		&\ba_{\rmr}\deh{-1}\left(\deh{-.5}\theta^{{\rm az}(\tau_n)}_\ell\deh{-1}-\deh{-.8}\varpi^{(\tau_{n})}, \theta^{{\rm el}(\tau_n)}_\ell\deh{-.5}\right)\deh{-.5}\ba_{\rmt}\deh{-1}\left(\deh{-.5}\phi^{{\rm az}(\tau_n)}_\ell\deh{-1}, \phi^{{\rm el}(\tau_n)}_\ell\deh{-.5}\right)^{\deh{-.5}*}\deh{-.8}\biggl)\deh{-.5},
	\end{align}
	where $d$ is the channel tap index, $T_\rms$ is the sampling interval, $t_{\rm off}^{(\tau_n)}$ is the unknown clock offset between the \ac{TX} and \ac{RX}, $f_{\rmp}(\cdot)$ is the filtering function that factors in filtering effects in the system, $\alpha_\ell^{(\tau_n)}$ and $t_\ell^{(\tau_n)}$ are the complex gain and the \ac{ToA} of the $\ell$-th path, $\ba_{\rmr}\left(\theta^{{\rm az}(\tau_n)}_\ell-\varpi^{(\tau_{n})}, \theta^{{\rm el}(\tau_n)}_\ell\right)$ represents the \ac{RX} array response evaluated at the azimuth and elevation \ac{AoA}, denoted as $\theta^{{\rm az}(\tau_n)}_\ell-\varpi^{(\tau_{n})}$ and $\theta^{{\rm el}(\tau_n)}_\ell$, and $\ba_{\rmt}\left(\phi^{{\rm az}(\tau_n)}_\ell, \phi^{{\rm el}(\tau_n)}_\ell\right)$ is the TX array response evaluated at the azimuth and elevation \ac{AoD}, denoted as $\phi^{{\rm az}(\tau_n)}_\ell$ and $\phi^{{\rm el}(\tau_n)}_\ell$. Note that, $\theta^{{\rm az}(\tau_n)}_\ell$ and $\theta^{{\rm el}(\tau_n)}_\ell$ are azimuth and elevation \ac{AoA}s in the global coordinate system, and the same applies to azimuth and elevation \ac{AoD}s $\phi^{{\rm az}(\tau_n)}_\ell$ and $\phi^{{\rm el}(\tau_n)}_\ell$. The array responses can be formulated in the Kronecker product form as 
	\begin{align}
		\begin{cases}\ba_\rmr(\theta^{\rm az}-\varpi,\theta^{\rm el})=\ba(\theta^\sprl,\theta^\bot)=\ba(\theta^\sprl)\otimes \ba(\theta^\bot)\\ 
			\ba_\rmt(\phi^{\rm az},\phi^{\rm el})=\ba(\phi^\sprl,\phi^\bot)=\ba(\phi^\sprl)\otimes \ba(\phi^\bot)\end{cases},
	\end{align}
	where $\theta^\sprl=\cos(\theta^{\rm el})\sin(\theta^{\rm az}-\varpi)$, $\theta^\bot=\sin(\theta^{\rm el})$, $\phi^\sprl=\cos(\phi^{\rm el})\sin(\phi^{\rm az})$, $\phi^\bot=\sin(\phi^{\rm el})$, and $\ba(\cdot)$ is the steering vector where $[\ba(\vartheta)]_n=e^{-j\pi(n-1)\vartheta}$ considering a half-wavelength element spacing for the planar arrays.
	
	Pilots in the form of $N_\rms\leq\min\{N^{\rm rf}_\rmt, N^{\rm rf}_\rmr\}$ data streams of length $Q$ are transmitted for channel tracking at each $\tau_n$ (we omit the upper right ``$(\tau_n)$" for simplicity for the following notation definition and notations),  where the $q$-th instance is denoted as $\bs[q]\in\bbC^{N_\rms\times 1}$ with $\bbE[\bs[q]\bs[q]^*]=\frac{1}{N_\rms}\bI_{N_\rms}$. Hybrid precoder and combiner are employed, which are denoted as $\bF =\bF^{\rm rf}\bF^{\rm bb}\in \bbC^{N_{\rmt}\times N_\rms}$ and $\bW =\bW^{\rm rf}\bW^{\rm bb}\in \bbC^{N_{\rmr}\times N_\rms}$, where $\bF^{\rm rf}$ and $\bF^{\rm bb}$ are the analog and digital precoders, and $\bW^{\rm rf}$ and $\bW^{\rm bb}$ are the analog and digital combiners. Within a channel tracking interval whose duration is less than the channel coherence time, $M$ precoder and combiner pairs are employed, denoted as $\bF_m$ and $\bW_m$, $m=1, 2,...,M$, for the $m$-th pair. Accordingly, the $q$-th instance of the received signal using $\bF_m$ and $\bW_m$ is given as
	\begin{equation}\label{receive_sig}
		\by_m[q]=\bW_m^*\sum\limits_{d=0}^{N_{\rmd}-1}\sqrt{P_{\rmt}}\bH_d\bF_m\bs[q-d]+\bW_m^*\bn_m[q],
	\end{equation}
	where $P_{\rmt}$ is the transmitted power, $N_\rmd$ is the number of channel taps, and $\bn_m[q]\sim \mathcal{N}({\bf 0},\frac{\sigma_{\bn}^2}{N_\rmr}\bI_{N_\rmr})$ is modeled as \ac{AWGN} where $\sigma^2_\bn=K_\rmB T_{\rmF}B_\rmc$, being $K_\rmB$ the Boltzmann constant and $T_\rmF$ the environmental temperature in Fahrenheit. Due to the noise being combined with $\bW^*_m$, it is no longer white. Therefore, we whiten the received signal as $\breve{\by}_m[q]=\bL_m^{-1}\by_m[q]$, where $\bL_m$ is computed via Cholesky decomposition of $\bW_m^*\bW_m=\bL_m\bL_m^*$, so that $\bbE\left[\bL_m^{-1}\bW_m^*\bn_m[q](\bL_m^{-1}\bW_m^*\bn_m[q])^*\right]=\sigma_\bn^2\bI_{N_\rms}$. Let $\breve{\bW}_m^*=\bL_m^{-1}\bW^*_m$ and $\breve{\bn}_m[q]=\bL_m^{-1}\bW_m^*\bn_m$ for simplicity, the whitened collected measurements can be written as
	\begin{equation}\label{receive_mat}
		\breve{\bY}_m =\breve{\bW}_m^* [\bH_0,...,\bH_{N_{\rmd}-1}]\left((\bI_{N_{\rmd}}\otimes\bF_m )\sqrt{P_\rmt}\bS \right)+\breve{\bN}_m,
	\end{equation}
	where $[\breve{\bY}_m]_{:,q}=\breve{\by}_m[q]$, $[\breve{\bN}_m]_{:,q}=\breve{\bn}_m[q]$, and $[\bS]_{:, q}=\left[\bs[q]; \bs[q-1];...;\bs[q-(N_{\rmd}-1)]\right]$.

	\section{Channel and Vehicle \ac{PO} tracking system}\label{sec:pos-track}
	This section provides detailed algorithms for joint channel and vehicle \ac{PO} tracking. The system diagram is shown in Fig. \ref{AlgPipeline}. We first introduce the F-MOMP algorithm, which accelerates the calculation of the product for the measurement and dictionary matrix and enhances computational efficiency compared to conventional OMP and MOMP algorithms, to realize mmWave channel tracking. Then VO-ChAT employing attention mechanisms predicts the current vehicle orientation based on the channel estimate sequence and orientation history. Following orientation compensation using the predicted orientation, the single-shot position estimate obtained by solving a \ac{WLS} problem is treated as the initial position estimate to be refined by VP-ChAT. Ultimately, VP-ChAT, the network inspired by the Transformer architecture, leverages historical channel estimate and position estimate sequences to provide the correction of the current single-shot position estimate, realizing precise vehicle position tracking.
	
	\vspace*{-2mm}
	\begin{figure*}[!t]
		\vspace{1ex}
		\hrule
		\vspace{1ex}
		{\small
			\begin{equation}\label{meaMat_all}
				\bUpsilon_m=\sqrt{P_\rmt}\begin{bmatrix} \bs[1]^\sfT[\bF_m^\sfT]_{:, 1}\bW_m^* & \hdots & \bs[1]^\sfT[\bF_m^\sfT]_{:, N_\rmt}\bW_m^* & \hdots &{\bf 0}^\sfT[\bF^\sfT]_{:, 1}\bW_m^* &\hdots & {\bf 0}^\sfT[\bF_m^\sfT]_{:, N_\rmt}\bW_m^* \\ 
					\vdots & \vdots & \vdots & \hdots & \vdots &\vdots &\vdots \\
					\bs[Q]^\sfT[\bF_m^\sfT]_{:, 1}\bW_m^* &\hdots & \bs[Q]^\sfT[\bF_m^\sfT]_{:, N_\rmt}\bW_m^* &\hdots &\bs[Q\deh{-1}-\deh{-1}(N_\rmd \deh{-1}-\deh{-1}1)]^\sfT[\bF_m^\sfT]_{:, 1}\bW_m^* & \hdots & \bs[Q\deh{-1}-\deh{-1}(N_\rmd\deh{-1} -\deh{-1}1)]^\sfT[\bF_m^\sfT]_{:, N_\rmt}\bW_m^*
				\end{bmatrix}.
		\end{equation}}
	\end{figure*}
	\begin{figure*}
		\vspace{1ex}
		\hrule
		\begin{align}\label{dicMat_all}
			[\bPsi]_{:, f_\bj}=\bigg[\left[\bp(\ddot{t}_{j_1})\right]_1 \left[\bar{\ba}(\ddot{\phi}_{j_2}^\sprl, \ddot{\phi}_{j_3}^\bot)\right]_1 &\ba(\ddot{\theta}_{j_4}^\sprl, \ddot{\theta}_{j_5}^\bot); \hdots; \left[\bp(\ddot{t}_{j_1})\right]_1 \left[\bar{\ba}(\ddot{\phi}_{j_2}^\sprl, \ddot{\phi}_{j_3}^\bot)\right]_{N_\rmt} \ba(\ddot{\theta}_{j_4}^\sprl, \ddot{\theta}_{j_5}^\bot);\hdots; \nonumber \\
			&\left[\bp(\ddot{t}_{j_1})\right]_{N_\rmd} \left[\bar{\ba}(\ddot{\phi}_{j_2}^\sprl, \ddot{\phi}_{j_3}^\bot)\right]_1 \ba(\ddot{\theta}_{j_4}^\sprl, \ddot{\theta}_{j_5}^\bot);\hdots; \left[\bp(\ddot{t}_{j_1})\right]_{N_\rmd} \left[\bar{\ba}(\ddot{\phi}_{j_2}^\sprl, \ddot{\phi}_{j_3}^\bot)\right]_{N_\rmt} \ba(\ddot{\theta}_{j_4}^\sprl, \ddot{\theta}_{j_5}^\bot)\bigg].
		\end{align}
	\end{figure*}
	\begin{figure*}
		\vspace{1ex}
		\hrule
		{\small
			\begin{align}\label{mult_element}
				[\bUpsilon_m]_{(q-1)N_\rms+1:qN_\rms, :}[\bPsi]_{:, f_\bj}&= 
				\sqrt{P_\rmt}\left(\sum\limits_{n_\rmd=1}^{N_\rmd} [\bp(\ddot{t}_{j_1})]_{n_\rmd}\bs[q\deh{-1}-\deh{-1}(n_\rmd\deh{-1}-\deh{-1}1)]^\sfT\right)\left(\sum\limits_{n_\rmt=1}^{N_\rmt}[\bF_m^\sfT]_{:, n_\rmt}\big[\bar{\ba}(\ddot{\phi}^\sprl_{j_2}, \ddot{\phi}^\bot_{j_3})\big]_{n_\rmt}\right) \left(\sum\limits_{n_\rmr=1}^{N_\rmr}[\bW_m^*]_{:,n_\rmr}\big[\ba(\ddot{\theta}^\sprl_{j_4}, \ddot{\theta}^\bot_{j_5})\big]_{n_\rmr}\right)\nonumber
				\\
				&=\Big[\sqrt{P_\rmt}\big[\bs[q],\bs[q-1],\hdots, {\bf 0}\big]\bp(\ddot{t}_{j_1})\Big]^\sfT 
				\left[\bF_m^\sfT\bar{\ba}(\ddot{\phi}^\sprl_{j_2}, \ddot{\phi}^\bot_{j_3})\right]\left[\bW_m^*\ba(\ddot{\theta}_{j_4}^\sprl, \ddot{\theta}_{j_5}^\bot)\right]\\
				&=\left[\zeta_q^\rmS(j_1)\right]^\sfT\deh{-1}\left[\zeta^\rmF_m(j_2,j_3)\right]\deh{-1}\left[\zeta^\rmW_m(j_4, j_5)\right]\in\bbC^{N_\rms\times 1}.
		\end{align}}
	\end{figure*}
	
	\subsection{F-MOMP Based Channel Tracking}\label{sec:chan-track}
	Before diving into the proposed F-MOMP channel tracking algorithm, we present a concise overview of the conventional OMP algorithm for channel estimation, followed by an explanation of how MOMP addresses the computational complexity issue of OMP through dimensional operations. Relevant notations are introduced throughout the discussion. Based on ${\rm vec}(\bA\bX\bB)=(\bB^\sfT\otimes \bA){\rm vec}(\bX)$, \eqref{receive_mat} can be written in the form
	\begin{align}
		{\rm vec}(\breve{\bY}_m) = \bUpsilon_m {\rm vec}([\bH_0, ...,\bH_{N_\rmd-1}])\nonumber+{\rm vec}(\breve{\bN}_m),
	\end{align}
	where $\bUpsilon_m=((\bI_{N_\rmd}\otimes\bF_m)\sqrt{P_\rmt}\bS)^\sfT\otimes\breve{\bW}^*_m\in\bbC^{QN_\rms\times N_\rmd N_\rmt N_\rmr}$ is the measurement matrix. The channel can be represented as ${\rm vec}([\bH_0,...,\bH_{N_\rmd-1}])=\bPsi\bx$ leveraging its sparsity, where $\bPsi$ is the dictionary formulated as
	\begin{align}\label{sing_Psi}
		\bPsi = \bA_\rmd\otimes(\overline{\bA}_\rmt\otimes \bA_\rmr)\in \bbC^{N_\rmr N_\rmt N_\rmd \times N_\rmr^\rma N_\rmt^\rma N_\rmd^\rma},
	\end{align}
	where $\bA_\rmd=\left[\bp(\ddot{t}_1), ...,\bp(\ddot{t}_{N^\rma_\rmd})\right]$ is the dictionary for the delay evaluated on the grid values $\{\ddot{t}_{j_1}|j_1=1,...,N^\rma_\rmd\}$, and $\bp(t)=[f_\rmp(0\cdot T_\rms-t),\dots,\allowbreak f_\rmp((N_\rmd-1)T_\rms-t)]^{\sfT}\in \bbR^{N_\rmd\times 1}$ is a sampled version of $f_\rmp(\cdot)$ mentioned in \eqref{chan_mod}; $\bA_\rmt=\bA_\rmt^\sprl\otimes\bA_\rmt^\bot$ is the  dictionary to evaluate azimuth and elevation \ac{AoD}s with $\bA_\rmt^\sprl=\left[\ba(\ddot{\phi}^\sprl_1),...,\ba(\ddot{\phi}^\sprl_{N^\rma_2})\right]\in \bbC^{N_\rmt^\rmx\times N^\rma_2}$ considering grids 
	$\left\{\ddot{\phi}^\sprl_{j_2}|j_2=1,...,N_2^\rma\right\}$ and $\bA_\rmt^\bot=\left[\ba(\ddot{\phi}^\bot_1),...,\ba(\ddot{\phi}^\bot_{N^\rma_3})\right]\in \bbC^{N_\rmt^\rmy\times N^\rma_3}$ considering grids $\{\ddot{\phi}^\bot_{j_3}|j_3=1,...,N_3^\rma\}$; and $\bA_\rmr=\bA_\rmr^\sprl\otimes\bA_\rmr^\bot$ is the dictionary to evaluate azimuth and elevation \ac{AoA}s, where $\bA_\rmr^\sprl\in\bbC^{N_\rmr^\rmx\times N_4^\rma}$ and $\bA_\rmr^\bot\in\bbC^{N_\rmr^\rmy\times N_5^\rma}$ are constructed similarly as $\bA_\rmt^\sprl$ and $\bA_\rmt^\bot$, respectively. Therefore, $N_1^\rma=N_\rmd^\rma$, $N_\rmt^\rma=N_2^\rma N_3^\rma$, and $N_\rmr^\rma=N_4^\rma N_5^\rma$. In addition, $\bx\in\bbC^{N_\rmr^\rma N_\rmt^\rma N_\rmd^\rma\times 1}$ is the sparse vector to be estimated the supports of which are the complex gains. To solve the following sparse recovery problem for channel estimation:
	\begin{align}
		\min\limits_{\bx}\left(\sum\limits_{m=1}^M\left\|{\rm vec}(\breve{\bY}_m)-\bUpsilon_m\bPsi\bx\right\|^2\right),
	\end{align}
	conventional OMP iteratively finds the supports of $\bx$, denoted as a set $\frak{x}=\{j\ |\ [\bx]_{j}\neq 0\},|\frak{x}|\leq N_{\rm est}$ where $N_{\rm est}$ is the number of channel components, based on peaks of the correlation with the residual calculated from the subspace projection. Once the supports are determined, the corresponding atoms in $\bPsi$, i.e., $\left\{[\bPsi]_{:, \ell_\rms}| \ell_\rms\in\frak{x}\right\}$, indicate the estimated delays and angles. The searching space size of the algorithm is $\prod_{k=1}^5N_k^\rma$, and with large antenna array and wide bandwidth for fine angular and delay domain resolutions, the resulted complexity $\cO\left(N_{\rm est}N_\rms Q\prod_{k=1}^5N_k^\rms N_k^\rma\right)$, where $N_1^\rms=N_\rmd$, $N_2^\rms=N_\rmt^\rmx$, $N_3^\rms=N_\rmt^\rmy$, $N_4^\rms=N_\rmr^\rmx$, and $N_5^\rms=N_\rmr^\rmy$, becomes prohibitive. To cope with the complexity issue, MOMP first formulates the problem with multidimensional operations:
	\begin{equation}\label{momp_opt}\small
		\min\limits_{\bX}\deh{-1}\left(\sum\limits_{m=1}^M\deh{-.5}\left\|{\rm vec}(\breve{\bY}_m)\deh{-1}-\deh{-1}\sum\limits_{\bi\in\frak{I}}\sum\limits_{\bj\in\frak{J}}[\bPhi_m]_{:, \bi}\left(\prod\limits_{k=1}^5[\bPsi_k]_{i_k,j_k}\deh{-1}\right)[\bX]_{\bj}\right\|^2\right),
	\end{equation}
	where $\bi=(i_1,...,i_5)\in\bbN_+^5$ and $\bj=(j_1,...,j_5)\in\bbN_+^5$ are multidimensional indices, and $\frak{I}=\left\{\bi|i_k=1,...,N_k^\rms\right\}$ and $\frak{J}=\left\{\bj|j_k=1,...,N_k^\rma\right\}$ are the index sets. The measurement tensor $\bPhi_m\in\bbC^{QN_\rms\otimes_{k=1}^5 N_k^\rms}$ relates to $\bUpsilon_m$ as $[\bPhi_m]_{:, \bi}=[\bUpsilon_m]_{:,f_{\bi}}$ where $f_\bi=\left(\sum_{k=1}^4(i_k-1)(\prod_{k'=k+1}^5N_{k'}^\rms)\right)+i_5$. The single dictionary $\bPsi$ calculated from the Kronecker product as in \eqref{sing_Psi} is separated into five independent dictionaries $\bPsi_k\in\bbC^{N_k^\rms\times N_k^\rma}$ for the five dimensions associated with delay, azimuth and elevation \ac{AoD}, and azimuth and elevation \ac{AoA}, i.e., $\bPsi_1=\bA_\rmd$, $\bPsi_2=\bA_\rmt^\sprl$, $\bPsi_3=\bA_\rmt^\bot$, $\bPsi_4=\bA_\rmr^\sprl$, and $\bPsi_5=\bA_\rmr^\bot$. Finally, the sparse vector $\bx$ becomes the sparse tensor $\bX\in \bbC^{\otimes_{k=1}^5 N_k^\rma}$ to be estimated, where $[\bX]_\bj=[\bx]_{f_\bj}$ with $f_\bj=\left(\sum_{k=1}^4(j_k-1)(\prod_{k'=k+1}^5 N_{k'}^\rma)\right)+j_5$. To solve \eqref{momp_opt}, alternating maximization is adopted to estimate the parameters for each dimension independently, with the cost of $N_{\rm iter}$ iterations to refine the estimates per dimension. Thus, the computational complexity is reduced to $\cO\left(N_{\rm est}N_\rms QN_{\rm iter}(\sum_{k=1}^5N_k^\rma)(\prod_{k=1}^5 N_k^{\rms})\right)$ compared with the conventional OMP, where the product term $\prod_{k=1}^5 N_k^\rma$ is transformed into a summation $N_{\rm iter}(\sum_{k=1}^5N_k^\rma)$. However, the complexity can still explode when employing large antenna arrays and wide bandwidth as in the product term $\prod_{k=1}^5N_k^\rms$. We hereby propose the F-MOMP algorithm that transforms the term into a summation for complexity reduction, while sticking with alternating maximization to determine the estimates for each dimension.
	
	The F-MOMP algorithm reduces the complexity of calculating the product of measurement and dictionary matrices. We first expand $\bUpsilon_m$ and $\bPsi$ as \eqref{meaMat_all} and \eqref{dicMat_all}, and based on the element-wise correspondence for the product operation, the product of $[\bUpsilon_m]_{(q-1)N_\rms+1:qN_\rms, :}$ and $[\bPsi]_{:,f_\bj}$ can be derived by factoring and then computing the product for each factor, as detailed in \eqref{mult_element}. Therefore, the sparse recovery problem can be written as
	\begin{equation}
		\scalebox{.96}{$
			\min\limits_{\bX}\deh{-.8}\sum\limits_{m=1}^M\deh{-.4}\sum\limits_{q=1}^Q\deh{-.5}\left\|\breve{\by}_m[q]\deh{-.8}-\deh{-1.2}\sum\limits_{\bj\in\frak{J}}\deh{-1}\left[\zeta_q^\rmS(j_1)\right]^\sfT\deh{-1}\left[\zeta^\rmF_m(j_2,j_3)\right]\deh{-1}\left[\zeta^\rmW_m(j_4, j_5)\right]\deh{-.8}\bX_\bj\right\|^2\deh{-1.2},$}
	\end{equation}
	where 
	\begin{align}
		&\zeta_q^\rmS(j_1)=\sqrt{P_\rmt}\big[\bs[q],\bs[q-1],\hdots, {\bf 0}\big]\bp(\ddot{t}_{j_1})\in \bbC^{N_\rms\times 1};\\
		&\zeta^\rmF_m(j_2,j_3)=\bF_m^\sfT\bar{\ba}(\ddot{\phi}^\sprl_{j_2}, \ddot{\phi}^\bot_{j_3})\in \bbC^{N_\rms\times 1};\\
		&\zeta^\rmW_m(j_4, j_5)=\bW_m^*\ba(\ddot{\theta}_{j_4}^\sprl, \ddot{\theta}_{j_5}^\bot)\in \bbC^{N_\rms\times 1},
	\end{align}
	and $\bX\in \bbC^{\otimes_{k=1}^5N_k^\rma}$ defined the same as that in \eqref{momp_opt} is the sparse tensor to be estimated with the MOMP algorithm. In the conventional MOMP, there is a step for estimate initialization for each dimension based on cost function approximation, then the alternating maximization algorithm is adopted to iteratively refine the estimates per dimension. However, in the channel tracking scenario, the channel estimates at time $\tau_{n-1}$ can be used as the estimate initialization at time $\tau_n$, i.e., the estimated support set at 
	$\tau_{n-1}$:
	\begin{align}\label{prev_sups}
		\hat{\frak{J}}^{(\tau_{n-1})}_{\rm sup}&=\left\{\hat{\bj}^{(\tau_{n-1})}_1,...,\hat{\bj}^{(\tau_{n-1})}_{N_{\rm est}}\right\},
	\end{align}
	where $\hat{\bj}_{n_{\rm est}}^{(\tau_{n-1})}=\deh{-.5}\left(\hat{j}_{1,n_{\rm est}}^{(\tau_{n-1})}\deh{-1},..., \hat{j}_{5,n_{\rm est}}^{(\tau_{n-1})}\right)$, provides the initialization for $\hat{j}_{k,n_{\rm est}}^{(\tau_n)}$ at time $\tau_n$ as $\hat{j}_{k,n_{\rm est}}^{(\tau_n)}\leftarrow \hat{j}_{k,n_{\rm est}}^{(\tau_{n-1})}$. In addition, the dictionaries at $\tau_n$ are constructed based on historical estimates as well. Let independent dictionaries $\bPsi_k$ defined previously be the full dictionaries used usually for initial access stages, we define the reduced dictionaries as $\bPsi_{k, n_{\rm est}}^{(\tau_n)}$ for the $n_{\rm est}$-th channel component at $\tau_n$, where the atoms from the full dictionaries corresponding to the previous estimates and their neighboring atoms are included:
	\begin{align}\label{reduc_dict_k}
		\bPsi_{k, n_{\rm est}}^{(\tau_n)}=[\bPsi_k]_{:,\hat{j}^{(\tau_{n-1})}_{k,n_{\rm est}}-g_k:\hat{j}^{(\tau_{n-1})}_{k,n_{\rm est}}+g_k},
	\end{align}
	where $g_k$ is the number of spanning grids for the neighboring atoms depending on the grid resolution of each $\bPsi_k$. Even with the reduced dictionaries, directly applying the OMP algorithm remains computationally intensive considering the high dictionary resolutions required for precise localization, especially when increasing $g_k$ for a larger searching space. Hence, we rely on the alternating maximization algorithm as in MOMP to iteratively estimate the support of each dimension for the $n_{\rm est}$-th channel component by solving the following optimization problem (the upper right time index ``$(\tau_n)$" is omitted for simplicity): 
	\begin{align}\label{iter_opt}
		&\max\limits_{j_{k,n_{\rm est}}}\ \sum\limits_{m=1}^M\frac{\left|\left(\bUpsilon_m[\bPsi]_{:, f_{\bj_{n_{\rm est}}}}\right)^*{\rm vec}(\breve{\bY}^{\rm res}_m)\right|}{\left\|\bUpsilon_m[\bPsi]_{:,f_{\bj_{n_{\rm est}}}}\right\|_2}\\
		&\text{ s.t. }\quad j_{k, n_{\rm est}}\in \frak{J}_{k, n_{\rm est}}, \ \bj_{n_{\rm est}}\notin\hat{\frak{J}}_{\rm sup}
	\end{align}
	where $\frak{J}_{k, n_{\rm est}}=\left\{\hat{j}_{k, n_{\rm est}}^{(\tau_{n-1})}-g_k,...,\hat{j}_{k, n_{\rm est}}^{(\tau_{n-1})}+g_k\right\}$, and $\breve{\bY}_m^{\rm res}$ --which is initialized using $\breve{\bY}_m$-- represents the residual after the subspace projection using the estimated supports. Specifically, in each optimization iteration $n_{\rm iter}\leq N_{\rm iter}$, the algorithm sequentially optimizes the estimate $\hat{j}_{k, n_{\rm est}}$ while fixing estimates of other dimensions $\hat{j}_{k',n_{\rm est}},\ k'\neq k$, until every $\hat{j}_{k, n_{\rm est}}$ is obtained. Thereafter, delay and angle estimates of each path are determined by indexing the grid values in the dictionaries using $\hat{\bj}_{n_{\rm est}}$. Finally, the estimated complex gain for each path $\hat{\alpha}_{n_{\rm est}}$ is acquired based on the estimated sparse vector as $\hat{\alpha}_{n_{\rm est}}=[\hat{\bx}]_{n_{\rm est}}$. The pseudo codes of the F-MOMP algorithm are presented in Algorithm \ref{algo_FMOMP}. The algorithm results in a complexity of $\cO\left(N_{\rm est}N_\rms QN_{\rm iter}(\sum_{k=1}^{5}N_k^\rma)(N_1^\rma+N_2^\rms N_3^\rms+N_4^\rms N_5^\rms)\right)$, which reduces the complexity by turning the multiplication term into the summation comparing with the MOMP \cite{palacios2022multidimensional, palacios2022low}, as specified in Table \ref{chanEstComplex}, while allows simultaneously estimating parameters across the five dimensions for delay, azimuth and elevation \ac{AoD}s, and azimuth and elevation \ac{AoA}s.
	We denote the estimated channel at $\tau_n$ containing $N_{\rm est}$ estimated paths without the compensation for the time-varying clock offset $t_{\rm off}^{(\tau_n)}$ and orientation $\varpi^{(\tau_{n})}$ as 
	\begin{align}\label{chan_with_offsets}
		\hat{\underline{\bZ}}_{\tau_n} = \left[\hat{\balpha}_{\tau_n}, \hat{\underline{\bt}}_{\tau_n}, \hat{\underline{\btheta}}^{\rm az}_{\tau_n}, \hat{\btheta}^{\rm el}_{\tau_n}, \hat{\bphi}^{\rm az}_{\tau_n}, \hat{\bphi}^{\rm el}_{\tau_n}\right]\in \bbR^{N_{\rm est}\times 6},
	\end{align} 
	where $\hat{\balpha}_{\tau_n} = \left[\left|\hat{\alpha}_1^{(\tau_n)}\right|,...,\left|\hat{\alpha}_{N_{\rm est}}^{(\tau_n)}\right|\right]^\sfT$ (the phase is irrelevant to acquire the position and orientation estimation \cite{chen2023learning}), $\hat{\underline{\bt}}_{\tau_n}=\left[\hat{t}^{(\tau_n)}_1\deh{-1}-\deh{-.8}\hat{t}_{\rm off}^{(\tau_n)},...,\hat{t}^{(\tau_n)}_{N_{\rm est}}\deh{-1}-\deh{-.8}\hat{t}_{\rm off}^{(\tau_n)}\right]^\sfT$,  $\hat{\underline{\btheta}}^{\rm az}_{\tau_n}=\left[\hat{\theta}^{{\rm az}(\tau_n)}_1\deh{-1}-\deh{-.8}\hat{\varpi}^{(\tau_n)}, ...,\hat{\theta}^{{\rm az}(\tau_n)}_{N_{\rm est}}\deh{-1}-\deh{-.8}\hat{\varpi}^{(\tau_n)}\right]^\sfT$,  $\hat{\btheta}^{\rm el}_{\tau_n}=\left[\hat{\theta}^{{\rm el}(\tau_n)}_1, ...,\hat{\theta}^{{\rm el}(\tau_n)}_{N_{\rm est}}\right]^\sfT$, $\hat{\bphi}^{\rm az}_{\tau_n}=\left[\hat{\phi}^{{\rm az}(\tau_n)}_1, ...,\hat{\phi}^{{\rm az}(\tau_n)}_{N_{\rm est}}\right]^\sfT$, and $\hat{\bphi}^{\rm el}_{\tau_n}=\left[\hat{\phi}^{{\rm el}(\tau_n)}_1, ...,\hat{\phi}^{{\rm el}(\tau_n)}_{N_{\rm est}}\right]^\sfT$.

	\begin{algorithm}[!t]  
		\caption{F-MOMP for channel tracking}  
		\label{algo_FMOMP}  
		{\small
			\begin{algorithmic}[1]   
				\State \textbf{Input:} 
				\Statex Vectorized received signals $\breve{\bf \bgamma}\deh{-1}\leftarrow\deh{-1}\left[{\rm vec}(\breve{\bY}^{(\tau_n)}_1);...;{\rm vec}(\breve{\bY}^{(\tau_n)}_M)\right]$; 
				\Statex Previous estimated supports $\hat{\frak{J}}_{\rm sup}^{(\tau_{n-1})}$ as in \eqref{prev_sups}; 
				\Statex The number of channel components $N_{\rm est}$;
				\State \textbf{Initialize:}
				\Statex The estimated support set $\hat{\frak{J}}_{\rm sup}^{(\tau_n)}\leftarrow \emptyset$;
				\Statex The subspace projection residual $\breve{\bgamma}^{\rm res}\leftarrow \breve{\bgamma}$;
				\For{$n_{\rm est}=1: N_{\rm est}$}
				\For{$k=1:5$}
				\State Initialize support estimates $\hat{j}_{k,n_{\rm est}}^{(\tau_n)}\leftarrow \hat{j}_{k,n_{\rm est}}^{(\tau_{n-1})}$;
				\State Construct reduced dictionaries $\bPsi_{k, n_{\rm est}}^{(\tau_n)}$ as in \eqref{reduc_dict_k}; 
				\State Form index sets $\frak{J}_k\leftarrow\left\{\hat{j}_{k,n_{\rm est}}^{(\tau_{n})}\deh{-1}-\deh{-.8}g_k,...,\hat{j}_{k,n_{\rm est}}^{(\tau_{n})}\deh{-1}+\deh{-.8}g_k\right\}$;
				\EndFor
				\State \textit{\% Factor calculation}
				\State $\bXi^\rmS_{j_1}\leftarrow\left[\zeta^\rmS_1(j_1)^\sfT;...;\zeta^\rmS_Q(j_1)^\sfT\right]$ for $j_1\in\frak{J}_1$;
				\State $\bXi^\rmF_{j_2,j_3}\leftarrow\left[\zeta^\rmF_1(j_2,j_3),...,\zeta^\rmF_M(j_2,j_3)\right]$ for $j_2\in\frak{J}_2$, $j_3\in\frak{J}_3$;
				\State $\bXi^\rmW_{j_4,j_5}\leftarrow\left[\zeta^\rmW_1(j_4,j_5),...,\zeta^\rmW_M(j_4,j_5)\right]$ for $j_4\deh{-.8}\in\frak{J}_4$, $j_5\in\frak{J}_5$;
				
				\For{$n_{\rm iter}=1:N_{\rm iter}$}
				\For{$k=1:5$}
				
				\State $\frak{J}\leftarrow\{\bj|j_k\in \frak{J}_k, j_{k'}=\hat{j}^{(\tau_n)}_{k',{\rm est}}, k'\neq k\} \backslash \hat{\frak{J}}_{\rm sup}^{(\tau_n)}$;
				\State $\bxi_{\bj}\deh{-1}\leftarrow{\rm vec}\left(\left(\bXi^\rmS_{j_1}\bXi^\rmF_{j_2, j_3}\right)\odot\bXi^\rmW_{j_4,j_5}\right)$, $\bj\in\frak{J}$, as in \eqref{mult_element};
				\State $\hat{j}_{k,n_{\rm est}}^{(\tau_n)}\leftarrow\arg\max\limits_{j_k}\frac{\left|\bxi_\bj^*\breve{\bgamma}^{\rm res}\right|}{\left\|\bxi_\bj\right\|_2}$ for solving \eqref{iter_opt};
				\EndFor
				\EndFor
				\State Collect support estimates $\hat{\bj}^{(\tau_n)}_{n_{\rm est}}=\left(\hat{j}_{1, n_{\rm est}}^{(\tau_n)},...,\hat{j}_{5, n_{\rm est}}^{(\tau_n)}\right)$;
				\State Retrieve channel parameters $\hat{t}^{(\tau_n)}_{n_{\rm est}}\deh{-1}=\deh{-.8}\ddot{t}_{\hat{j}_{1,n_{\rm est}}^{(\tau_n)}}\deh{-1}$, $\hat{\phi}^{\sprl(\tau_n)}_{n_{\rm est}}\deh{-1}=\deh{-.8}\ddot{\phi}^\sprl_{\hat{j}^{(\tau_n)}_{2, n_{\rm est}}}\deh{-1}$,  $\hat{\phi}^{\bot(\tau_n)}_{n_{\rm est}}\deh{-1}=\deh{-.8}\ddot{\phi}^\bot_{\hat{j}^{(\tau_n)}_{3, n_{\rm est}}}\deh{-1}$, $\hat{\theta}^{\sprl(\tau_n)}_{n_{\rm est}}\deh{-1}=\deh{-.8}\ddot{\theta}^\sprl_{\hat{j}^{(\tau_n)}_{4, n_{\rm est}}}\deh{-1}$, and $\hat{\theta}^{\bot(\tau_n)}_{n_{\rm est}}\deh{-1}=\deh{-.8}\ddot{\theta}^\bot_{\hat{j}^{(\tau_n)}_{5, n_{\rm est}}}\deh{-1}$; 
				\State Update support set $\hat{\frak{J}}_{\rm sup}^{(\tau_n)}\leftarrow\hat{\frak{J}}_{\rm sup}^{(\tau_n)}\cup\left\{\hat{\bj}^{(\tau_n)}_{n_{\rm est}}\right\}$;
				\State \textit{\% Subspace projection and residual update}
				\State $\hat{\bx} \leftarrow\left[\bxi_{\hat{\bj}_1^{(\tau_n)}},...,\bxi_{\hat{\bj}^{(\tau_n)}_{n_{\rm est}}}\right]^\dag \breve{\bgamma}$;
				
				\State $\breve{\bgamma}^{\rm res} \leftarrow \breve{\bgamma}-\left[\bxi_{\hat{\bj}_1^{(\tau_n)}},...,\bxi_{\hat{\bj}^{(\tau_n)}_{n_{\rm est}}}\right]\hat{\bx}$;
				\EndFor
				\State Retrieve path complex gains where $\hat{\alpha}_{n_{\rm est}} = [\hat{\bx}]_{n_{\rm est}}$;
				\State \textbf{Output:} $\hat{\frak{J}}_{\rm sup}^{(\tau_n)}$ and estimated channel parameters for each path.

			\end{algorithmic}
		}
	\end{algorithm}
	\begin{table}[!t]\centering
		\resizebox{\linewidth}{!}{
			\begin{tabular}{m{.45\linewidth}l}
				\arrayrulecolor{black}\toprule
				\multicolumn{1}{c}{Method}             & \multicolumn{1}{c}{Complexity}                                                                                              \\ \midrule
				Conventional OMP \cite{venugopal2017channel, rodriguez2018frequency} & $\cO\left(N_{\rm est}N_\rms Q\prod\limits_{k=1}^5N_k^\rms N_k^\rma\right)$                 \\ \arrayrulecolor{lightgray}\hdashline
				MOMP \cite{palacios2022multidimensional, palacios2022low}                   & $\cO\left(N_{\rm est}N_\rms QN_{\rm iter}(\sum\limits_{k=1}^5N_k^\rma)(\prod\limits_{k=1}^5 N_k^{\rms})\right)$                                                       \\ \arrayrulecolor{lightgray}\hdashline
				
				\textbf{F-MOMP (proposed)} & $\cO\left(N_{\rm est}N_\rms QN_{\rm iter}(\sum\limits_{k=1}^{5}N_k^\rma)(N_\rmd+N_2^\rms N_3^\rms+N_4^\rms N_5^\rms)\right)$ \\
				\arrayrulecolor{black}\bottomrule
			\end{tabular}
		}
		\caption{Complexity comparisons for various channel estimation algorithms. }
		\label{chanEstComplex}
	\end{table}
	\begin{figure*}[!t]
		\centering
		\includegraphics[width=\linewidth]{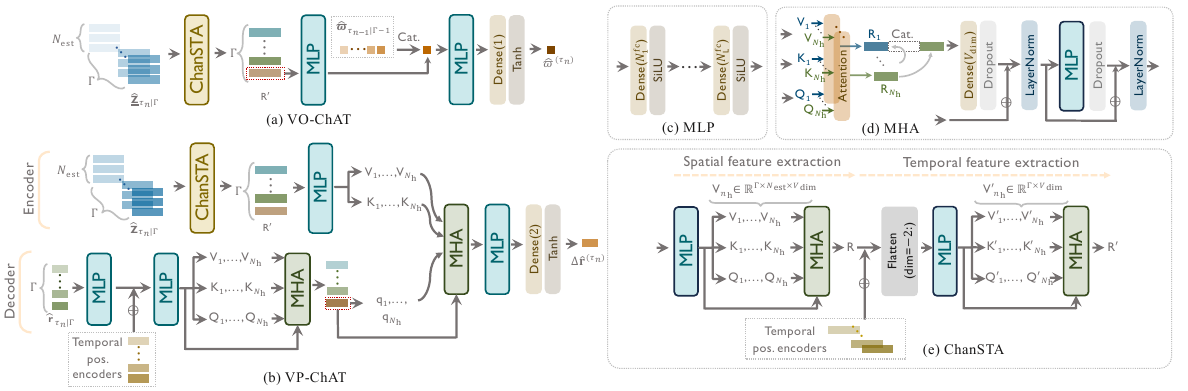}
		\caption{(a) VO-ChAT architecture to track the vehicle orientations; (b) VP-ChAT architecture to realize information exchange between the estimated channel sequence and the vehicle trajectory for corrections of the single-shot position estimates; (c) \ac{MLP} layer constructions; (d) \ac{MHA} module design; (e) The ChanSTA module to realize first spatial feature extraction and then temporal feature extraction for the estimated channel sequence.}
		\label{VPO-ChAT_Arch}
	\end{figure*}
	
	\subsection{VO-ChAT for Vehicle Orientation Tracking}
	The absence of orientation knowledge limits the vehicle's estimated angles to relative values w.r.t the planar array rather than the absolute values in the global coordinate system. Consequently, precisely determining the vehicle's position becomes infeasible. While orientation changes --influenced by multiple environmental factors including winds, turbulence, and atmospheric disturbances \cite{van2023modeling}-- are challenging to model accurately, we turn to \ac{DL}, specifically, the attention schemes, that have been broadly studied in prior work to address context-aware problems \cite{de2022attention}, and propose VO-ChAT as illustrated in Fig. \ref{VPO-ChAT_Arch}a that takes in the estimated channel sequence and previous orientation estimates to infer the current orientation
	\begin{equation}
		\hat{\varpi}^{(\tau_n)} = {\rm VO}\text{-}{\rm ChAT}\left(\hat{\underline{\bsfZ}}_{\tau_n|\Gamma}, \hat{\bvarpi}_{\tau_{n-1}|\Gamma-1}; \bsfW^\rmo\right),
	\end{equation}
	where $\Gamma$ is the length of history information to be considered,  $\hat{\underline{\bsfZ}}_{\tau_n|\Gamma}={\rm stack}\left(\hat{\underline{\bZ}}_{\tau_{n-\Gamma+1}},...,\hat{\underline{\bZ}}_{\tau_n}\right)\in\bbR^{\Gamma\times N_{\rm est}\times 6}$ is the estimated channel sequence from time $\tau_{n-\Gamma+1}$ to $\tau_n$, $\hat{\bvarpi}_{\tau_{n-1}|\Gamma-1}=\left[\hat{\varpi}^{(\tau_{n-\Gamma+1})},...,\hat{\varpi}^{(\tau_{n-1})}\right]\in \bbR^{\Gamma-1}$ is the vector containing previous determined orientations from $\tau_{n-\Gamma+1}$ to $\tau_{n-1}$, and $\bsfW^\rmo$ represents the learnable network matrices.

	VO-ChAT firstly processes $\hat{\underline{\bsfZ}}_{\tau_n|\Gamma}$ with a \ac{ChanSTA} module depicted in Fig. \ref{VPO-ChAT_Arch}e, which is composed of two self \ac{MHA} blocks (Fig. \ref{VPO-ChAT_Arch}d), one for extracting spatial features w.r.t the estimated $N_{\rm est}$ paths at each time step, and the other one for analyzing the temporal channel evolution features for the input channel sequence of length $\Gamma$. Specifically, $\hat{\underline{\bsfZ}}_{\tau_n|\Gamma}$ goes through a \ac{MLP} module consisting of dense layers and activation layers to obtain three types of abstract representations: \textit{Value} $\{\bsfV_1,...,\bsfV_{N_\rmh} \}$ with $\bsfV_{n_\rmh}\in \bbR^{\Gamma\times N_{\rm est}\times V_{\rm dim}}$, \textit{Key} $\{\bsfK_1,...,\bsfK_{N_\rmh}\}$ with $\bsfK_{n_\rmh}\in\bbR^{\Gamma\times N_{\rm est}\times K_{\rm dim}}$, and \textit{Query} $\{\bsfQ_1,...,\bsfQ_{N_\rmh}\}$ with $\bsfQ_{n_\rmh}\in\bbR^{\Gamma\times N_{\rm est}\times Q_{\rm dim}}$, where $V_{\rm dim}$, $K_{\rm dim}$, and $Q_{\rm dim}=K_{\rm dim}$ are the embedding dimensions for each type of the representations, and $N_\rmh$ is the number of attention heads in the \ac{MHA} mechanism. The attention operation for the $n_\rmh$-th head to extract path-wise spatial features is mathematically formulated as 
	\begin{align}\label{AttEqu}
		[\bsfR_{n_\rmh}]_{\sfi,\sfk, :}&={\rm Attention}\deh{-1}\left(\left[\bsfQ_{n_\rmh}\right]_{\sfi,\sfk, :}\deh{-1},\left[\bsfK_{n_\rmh}\right]_{\sfi,:, :}\deh{-1},\left[\bsfV_{n_\rmh}\right]_{\sfi,:, :}\right)\\
		&={\rm Softmax}\left(\frac{[\bsfQ_{n_\rmh}]_{\sfi,\sfk, :,  }\left[\bsfK_{n_\rmh}\right]_{\sfi,:,:}^{\sfT}}{\sqrt{K_{\rm dim}}}\right)\left[\bsfV_{n_\rmh}\right]_{\sfi,:, :}.
	\end{align}
	Here, $\bsfR_{n_\rmh}$ has the same shape as $\bsfV_{n_\rmh}$, and each row of $[\bsfR_{n_\rmh}]_{\sfi,:, :}$ corresponds to an estimated channel path factoring in other paths' information within the same estimation time frame, i.e., paths with higher estimation confidence at each time step should be prioritized for subsequent processing. As shown in Fig. \ref{VPO-ChAT_Arch}d, outputs from individual attention heads are concatenated and processed through a dense layer resulting in a dimension of $V_{\rm dim}$, and then connected to the normalization layers to ensure consistent feature scaling and effective information propagation through the network. Let $\bsfR\in\bbR^{\Gamma\times N_{\rm est}\times V_{\rm dim}}$ represent the output from the spatial feature extraction block, position encoding is then applied to $\bsfR$ to preserve the chronological order of the estimated channels before temporal feature extraction. The resulting tensor is flattened along the last two dimensions with the out shape of $\Gamma\times N_{\rm est}V_{\rm dim}$. Subsequent processing through a \ac{MLP} module generates three types of channel representations, $\{\bsfV'_1,...,\bsfV'_{N_\rmh} \}$ with $\bsfV'_{n_\rmh}\in \bbR^{\Gamma\times V_{\rm dim}}$, $\{\bsfK'_1,...,\bsfK'_{N_\rmh}\}$ with $\bsfK'_{n_\rmh}\in\bbR^{\Gamma\times K_{\rm dim}}$, and $\{\bsfQ'_1,...,\bsfQ'_{N_\rmh}\}$ with $\bsfQ'_{n_\rmh}\in\bbR^{\Gamma\times Q_{\rm dim}}$, where each row of the representation matrices corresponds to information from a specific time step. Thereafter, the three types of representations are processed through the \ac{MHA} module where the attention operation for the $n_\rmh$-th attention head becomes $[\bsfR'_{n_\rmh}]_{\sfi,:}={\rm Attention}\left([\bsfQ'_{n_\rmh}]_{\sfi, :}, \bsfK'_{n_\rmh}, \bsfV'_{n_\rmh}\right)$. The resulting output passes through a dense layer and is residually connected to the normalization layers,  similar to the structure of the preceding spatial feature extraction block. The output from the temporal feature extraction block is the representation $\bsfR'\in \bbR^{\Gamma\times V_{\rm dim}}$ which emphasizes more accurately estimated channels within the sequence and incorporates temporal evolution features. To acquire the orientation estimate at the current time step, the final row of $\bsfR'$, i.e., $[\bsfR']_{-1, :}$ indicating the current information, is selected to go through \ac{MLP} layers, along with the concatenated previous orientation estimates $\hat{\bvarpi}_{\tau_{n-1}|\Gamma-1}$, to produce the current orientation estimate $\hat{\varpi}^{(\tau_n)}$. Notably, the channel estimates provide essential information about the propagation environment, and enable the network to adjust and mitigate sequential orientation prediction errors. In summary, this approach achieves orientation tracking by incorporating channel and orientation histories, leveraging the temporal consistency between channel variations and orientation changes, and capturing the inherent relationship between vehicle motion and channel evolution.  
	
	\subsection{Geometric Transformation for Single-Shot Localization}
	Once the vehicle orientation $\varpi^{(\tau_{n})}$ is determined, the estimated relative azimuth \ac{AoA}s can be compensated to acquire the angles in the global coordinate system as $\hat{\btheta}_{\tau_n}^{\rm az}=\hat{\underline{\btheta}}_{\tau_n}^{\rm az}+\hat{\varpi}^{(\tau_n)}=\left[\hat{\theta}_1^{{\rm az}(\tau_n)},...,\hat{\theta}_{N_{\rm est}}^{{\rm az}(\tau_n)}\right]^\sfT$. To derive the vehicle's position, we first leverage the concepts of \ac{DoD} and \ac{DoA} in the form of unitary vectors, denoted as $\vv{\upvarphi_\ell}=[\cos(\phi_\ell^{\rm el})\cos(\phi_\ell^{\rm az}), \cos(\phi_\ell^{\rm el})\sin(\phi_\ell^{\rm az}), \sin(\phi_\ell^{\rm el})]^\sfT$ and $\vv{\upvartheta_\ell}=[\cos(\theta_\ell^{\rm el})\cos(\theta_\ell^{\rm az}), \cos(\theta_\ell^{\rm el})\sin(\theta_\ell^{\rm az}), \sin(\theta_\ell^{\rm el})]^\sfT$, and formulate the geometric relationship between the BS and the vehicle for each first order reflection $\ell$  satisfying
	\begin{align}
		&\br^{(\tau_n)}_\rmv+d_\ell^{\upvartheta(\tau_n)}\cdot\vv{\upvartheta_\ell}^{(\tau_n)} = \br_\rmB+d_\ell^{\upvarphi(\tau_n)}\cdot\vv{\upvarphi_\ell}^{(\tau_n)};\label{geo_formu}\\ &d_\ell^{\upvartheta(\tau_n)}+d_\ell^{\upvarphi(\tau_n)}=\left(t^{(\tau_n)}_\ell+t^{(\tau_n)}_{\rm off}\right)\cdot v_\rmc,\label{time_formu}
	\end{align}
	where $\br^{(\tau_n)}_\rmv$ is the vehicle's 3D position at $\tau_n$, $\br_\rmB$ is the known array position on the BS, $d_\ell^{\upvartheta(\tau_n)}$ is the distance between the vehicle and the scattering point, $d_\ell^{\upvarphi(\tau_n)}$ is the distance between the scattering point and the BS, and $v_\rmc$ is the light speed. Note that \eqref{geo_formu} and \eqref{time_formu} hold for \ac{LOS} situations as well assuming a pseudo scattering point in the middle of the \ac{LOS} path. Before resolving \eqref{geo_formu} and \eqref{time_formu} to determine the vehicle's position, it is imperative to identify and select the \ac{LOS}/first order reflections, as it allows for the exclusion of higher order \ac{MPCs} that will introduce errors in the following localization process. While our previous work \cite{chen2023learning} employs a neural network for path order classification, we leverage the tracking scenario's inherent advantages here. Specifically, we assume the height of the vehicle array is known and remained consistent along a trajectory as $\left[\br_\rmv^{(\tau_n)}\right]_3=h_\rmv$, and substitute $d_\ell^{\upvarphi(\tau_n)}=\left(h_\rmv+d_\ell^{\upvartheta(\tau_n)}\cdot\left[\vv{\upvartheta_\ell}^{(\tau_n)}\right]_3-[\br_\rmB]_3\right)/\left[\vv{\upvarphi_\ell}^{(\tau_n)}\right]_3$ into \eqref{time_formu} to derive $\hat{d}_\ell^{\upvartheta(\tau_n)}=\frac{\left[\hat{\vv{\upvarphi_\ell}}^{(\tau_n)}\right]_3\cdot\left(\hat{t}^{(\tau_n)}_\ell+\hat{t}^{(\tau_0)}_{\rm off}\right)\cdot v_\rmc+[\br_\rmB]_3-h_\rmv}{\left[\hat{\vv{\upvarphi_\ell}}^{(\tau_n)}\right]_3+\big[\hat{\vv{\upvartheta_\ell}}^{(\tau_n)}\big]_3}$, where $\hat{t}_{\rm off}^{(\tau_0)}$ is the clock offset estimated during the initial access stage \cite{chen2023learning} and the subsequent time-varying clock offsets $t_{\rm off}^{(\tau_n)}$ should be attributed to small drifts. Then path $\ell$ is discarded for localization if $\hat{d}_\ell^{\upvartheta(\tau_n)}\leq[\br_\rmB]_3$. In addition, the estimated path gain should be above a threshold to guarantee the channel tracking accuracy, i.e., an estimated path is also discarded if $|\hat{\alpha}_\ell|\leq |\alpha_{\rm th}|$. Afterwards, for all the selected paths $\ell\in \frak{L}$, where $\frak{L}$ is the set containing the estimated LOS and/or first order reflections, we substitute $d_\ell^{\upvarphi(\tau_n)}=v_\rmc t_\ell^{(\tau_n)}+v_\rmc t_{\rm off}^{(\tau_n)}-d_\ell^{\upvartheta(\tau_n)}$ into \eqref{geo_formu} as
	\begin{align}\label{geo_formu2}
		\br_\rmv^{(\tau_n)}+d_\ell^{\vartheta(\tau_n)}&\left(\vv{\upvartheta_\ell}^{(\tau_n)}+\vv{\upvarphi_\ell}^{(\tau_n)}\right)-v_\rmc t_{\rm off}^{(\tau_n)}\vv{\upvarphi_\ell}^{(\tau_n)}\nonumber\\
		&=\br_\rmB+v_\rmc t_\ell^{(\tau_n)}\vv{\upvarphi_\ell}^{(\tau_n)}.
	\end{align}
	Therefore, a \ac{WLS} estimation problem can be formulated:
	\begin{align}\label{loc_WLS}
		\begin{bmatrix}
			w_1\bB_1\deh{-0.5} \\ \vdots \\w_{|\frak{L}|}\bB_{|\frak{L}|}
		\end{bmatrix}\bo=\begin{bmatrix}
			w_1\bb_1 \\ \vdots \\w_{|\frak{L}|}\bb_{|\frak{L}|}
		\end{bmatrix},
	\end{align}
	where $w_\ell$ is the weight assigned to path $\ell$ proportional to its estimated gain $|\hat{\alpha}_\ell|$ in decibel, $\bB_\ell=\left[\bB'_\ell, \bB^{''}_\ell\right]\in \bbR^{3\times (3+|\frak{L}|)}$ with $\bB'_\ell=\begin{bmatrix}
		\begin{aligned}\bI_2\\ {\bf 0}_{1\times2}\end{aligned}& -\hat{\vv{\upvarphi_\ell}}^{(\tau_n)}
	\end{bmatrix}$ and $\bB^{''}_\ell\in\bbR^{3\times |\frak{L}|}$ containing columns of zeros except its $\ell$-th column given by $\left[\bB^{''}_\ell\right]_{:, \ell}=\hat{\vv{\upvartheta_\ell}}^{(\tau_n)}+\hat{\vv{\upvarphi_\ell}}^{(\tau_n)}$, $\bb_\ell=\br_\rmB-[0, 0,h_\rmv]^\sfT+v_\rmc \hat{t}_\ell^{(\tau_n)}\hat{\vv{\upvarphi_\ell}}^{(\tau_n)}$, and the vector containing the unknown variables to be estimated with the LS estimation algorithm is defined as 
	\begin{align}
		\bo=\left[[\br_\rmv^{(\tau_n)}]_{1:2}^\sfT,v_\rmc t_{\rm off}^{(\tau_n)},d_1^{\upvartheta(\tau_n)}\deh{-1},...,d^{\upvartheta(\tau_n)}_{|\frak{L}|}\right]^\sfT.
	\end{align} 
	By solving \eqref{loc_WLS}, the single-shot 2D localization result is given by $\hat{\br}_{\rmv_\sprl}^{(\tau_n)}=[\hat{\bo}]_{1:2}$, the clock offset is determined as $\hat{t}_{\rm off}^{(\tau_n)}=\frac{[\hat{\bo}]_3}{v_\rmc}$, and the estimated absolute \ac{ToA}s are accordingly obtained as $\hat{\bt}_{\tau_n}=\hat{\underline{\bt}}_{\tau_n}+\hat{t}_{\rm off}^{(\tau_n)}=\left[\hat{t}_1^{(\tau_n)},...,\hat{t}_{N_{\rm est}}^{(\tau_n)}\right]$. 
	
	We denote $\hat{\bsfZ}_{\tau_n|\Gamma}={\rm stack}\left(\hat{\bZ}_{\tau_{n-\Gamma+1}},...,\hat{\bZ}_{\tau_n}\right)$ for the subsequent position tracking task, where $\hat{\bZ}_{\tau_n} = \left[\hat{\balpha}_{\tau_n}, \hat{\bt}_{\tau_n}, \hat{\btheta}^{\rm az}_{\tau_n}, \hat{\btheta}^{\rm el}_{\tau_n}, \hat{\bphi}^{\rm az}_{\tau_n}, \hat{\bphi}^{\rm el}_{\tau_n}\right]\in \bbR^{N_{\rm est}\times 6}$ with the time offset and orientation compensated should be distinguished from $\hat{\underline{\bZ}}_{\tau_n}$ defined in \eqref{chan_with_offsets}.

	\subsection{VP-ChAT for Vehicle Position Tracking}
	While solving \eqref{loc_WLS} yields the single-shot localization results, incorporating historical trajectory information and calibrating  $\hat{\br}_{\rmv_\sprl}^{(\tau_n)}$ is beneficial to enhance the accuracy. To this end, we propose a second network VP-ChAT built upon the architecture of VO-ChAT, as illustrated in Fig. \ref{VPO-ChAT_Arch}b. In VP-ChAT, the ChanSTA module --structurally identical to that of VO-ChAT-- together with an additional \ac{MLP} module serves as an encoder, which captures the complex multipath characteristics and their temporal evolution.  Concurrently, a decoder processes the trajectory information within the given time period, then generates the \textit{query} representations of the position information to perform cross \ac{MHA} with the encoder outputs to request for a correction of the current single-shot position estimate, i.e., 
	\begin{align}
		\Delta\hat{\br}^{(\tau_n)}={\rm VP}\text{-}{\rm ChAT}\left(\hat{\bsfZ}_{\tau_n|\Gamma}, \hat{\bsfr}_{\tau_n|\Gamma};\bsfW^\rmp\right),
	\end{align}
	where $\Delta\hat{\br}^{(\tau_n)}$ is the correction vector for $\hat{\br}_{\rmv_\sprl}^{(\tau_n)}$ so that the corrected position is $\widetilde{\br}_{\rmv_\sprl}^{(\tau_n)}=\hat{\br}_{\rmv_\sprl}^{(\tau_n)}+\Delta\hat{\br}^{(\tau_n)}$, $\hat{\bsfZ}_{\tau_n|\Gamma}$ defined previously --the channel sequence with the orientation and clock offset compensated-- serves as the encoder input, $\hat{\bsfr}_{\tau_n|\Gamma}$ is the decoder input comprising the historical corrected position estimates $\widetilde{\br}_{\rmv_\sprl}^{(\tau_{n'})}$ for $n'=n-\Gamma+1,...,n-1$ and the current single-shot position estimate $\hat{\br}_{\rmv_\sprl}^{(\tau_n)}$, denoted as $\hat{\bsfr}_{\tau_n|\Gamma}=\left[\widetilde{\br}^{(\tau_{n-\Gamma+1})}_{\rmv_\sprl};...;\widetilde{\br}^{(\tau_{n-1})}_{\rmv_\sprl};\hat{\br}_{\rmv_\sprl}^{(\tau_n)}\right]\in\bbR^{\Gamma\times 2}$, and $\bsfW^\rmp$ is the learnable network parameters. In detail, the encoder extracts the spatial and temporal evolution features of the estimated channels similar to that in the orientation tracking task, and generates $N_\rmh$ pairs of \textit{value} and \textit{key} representations of the estimated channel sequence, denoted as $\{\bsfV_1,...,\bsfV_{N_\rmh}\}$ with $\bsfV_{n_\rmh}\in \bbR^{\Gamma\times V_{\rm dim}}$ and $\{\bsfK_1,...,\bsfK_{N_\rmh}\}$ with $\bsfK_{n_\rmh}\in \bbR^{\Gamma\times K_{\rm dim}}$ for the following cross attention operations. At the decoder, $\hat{\bsfr}_{\tau_n|\Gamma}$ is processed by \ac{MLP} layers for feature expansion, followed by the positional encoding to preserve chronicle order. The resulting tensor is processed to generate the three types of representations for self \ac{MHA} operations, which extracts the vehicle's moving patterns and produces a sequence of dimension $\Gamma\times V_{\rm dim}$, where each row represents the position information at a specific time step while incorporating contextual information from the entire input trajectory sequence. To facilitate current position correction, the final row of the sequence indicating the information from the current time step is transformed into $N_\rmh$ \textit{query} vectors, denoted as $\{\bsfq_1,...,\bsfq_{N_\rmh}\}$, to perform cross \ac{MHA} with the encoder outputs, i.e., ${\rm Attention}\left(\bsfq_{n_\rmh},\bsfK_{n_\rmh},\bsfV_{n_\rmh}\right)$ for $n_\rmh=1,...,N_\rmh$. Finally, the output from the cross \ac{MHA} mechanism undergoes additional processing through \ac{MLP} layers and yields the current position correction vector. In summary, the encoder-decoder architecture of VP-ChAT maintains temporal coherence for the channel and trajectory sequences, and the cross attention mechanisms establish the connections among the channel evolution, the vehicle’s trajectory, and system errors introduced by the channel estimation and localization methods, hereby achieving precise position refinement.
	
	\section{Simulation Results}\label{sec:SimResults}
	\begin{figure}[!t]
		\centering
		\includegraphics[width=.7\linewidth]{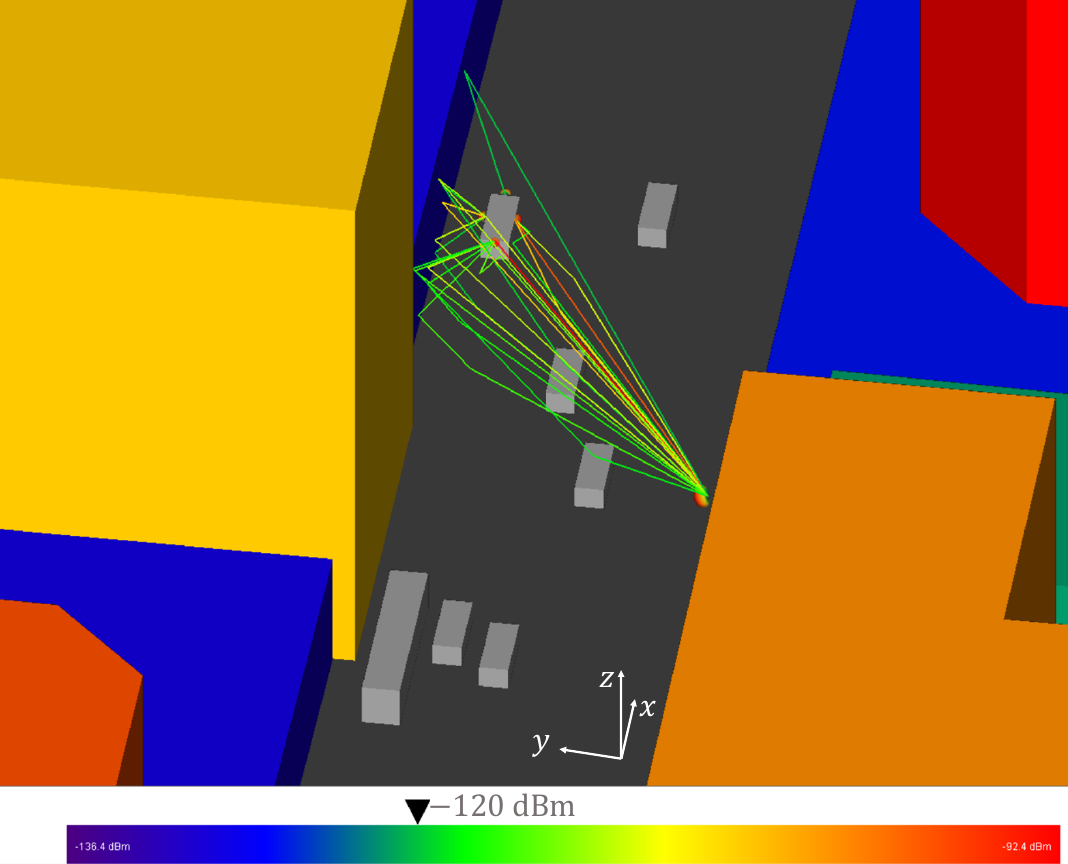}
		\caption{Ray-tracing simulation example in the urban canyon environment. The \ac{MPCs} with gains $\geq -120$ dBm are plotted.} 
		\label{Sim_env}
	\end{figure}
	This section presents the mmWave vehicular system setups, followed by the analysis of the experimental results. We first present the channel tracking performance using the F-MOMP algorithm to demonstrate its accuracy for vehicle localization. Subsequently, we present the orientation prediction results using VO-ChAT and evaluate the vehicle position tracking performance using VP-ChAT after orientation compensation, comparing these results with \ac{SOTA} localization methods with mmWave communication channels. 
	
	As the ray-tracing simulation snapshot depicted in Fig. \ref{Sim_env}, we consider an urban canyon environment within a rectangular cuboid with opposite vertices at points $[-13, -123, 0]$ (m) and $[231, 85, 56]$ (m). The environmental configurations including the surface materials follow the settings in \cite{anum2020passive}. The cars and trucks are distributed across four lanes in the center according to the \ac{3GPP} standard technical report \cite{3GPPVehEnv}, and move at the speed limits assigned to each lane: $60$, $50$, $25$, and $15$ km/h. We pick an active vehicle driving at $60$ km/h on the first lane for the tracking experiment, with its orientation dynamically adjusted according to the driver behavior model by setting $\br^\star({\tau+T_{\rm la}})$ on the lane centerline with looking ahead time $T_{\rm la}=0.5$ s, driver gain $K_\rme=2$, leading time constant $T_\rme=0.2$, the mean and variance of the wheel steering rate $\dot{\omega}=1.3$ rad/s and $\sigma_\omega^2=0.17^2$. Ray-tracing simulations are conducted at a carrier frequency of $f_\rmc=73$ GHz with the snapshots captured at $\Delta\tau=10$ ms intervals until the active vehicle reaches the lane end. We generate 32 trajectories where the vehicles start from randomly selected positions on each lane, each of which contains simulation results of $\sim 250$ snapshots.	
	\begin{figure*}
		\centering
		\begin{minipage}[!t]{0.32\textwidth}
			\centering
			\includegraphics[width=\textwidth]{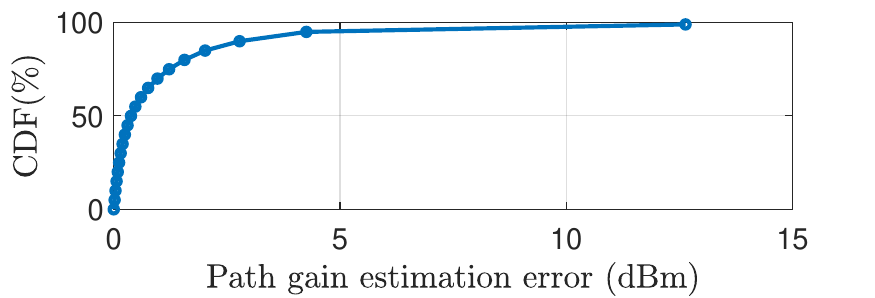}
			\text{(a) Channel gain estimates}\\
			\includegraphics[width=\textwidth]{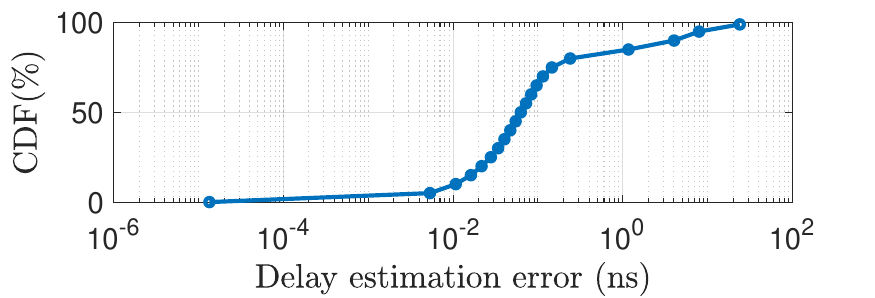}\label{chan_track_err_toa}
			\text{(b) \ac{ToA} estimates}
		\end{minipage}
		\hfill
		\begin{minipage}[!t]{0.32\textwidth}
			\centering
			\includegraphics[width=\textwidth]{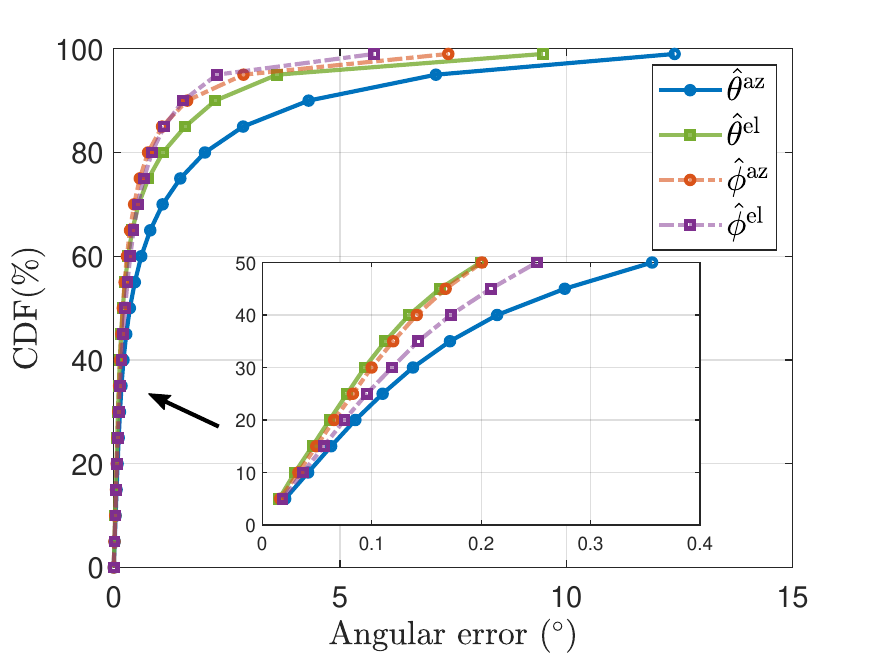}
			\text{(c) Angle estimates}
		\end{minipage}
		\hfill
		\begin{minipage}[!t]{0.32\textwidth}
			\centering
			\includegraphics[width=\textwidth]{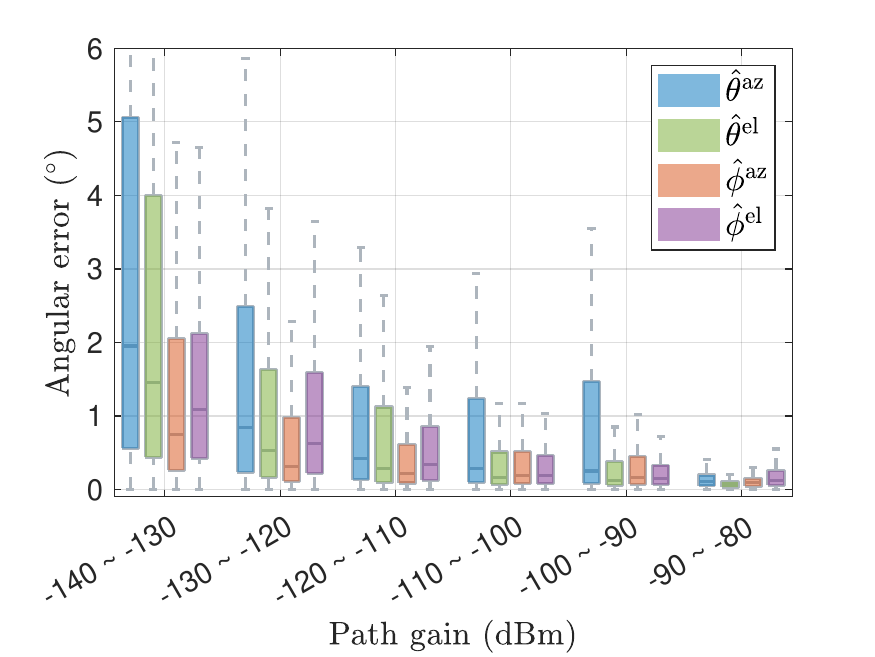}
			\text{(d) Angle estimates w.r.t path gains}
		\end{minipage}
		\caption{Channel tracking performance with the fixed transmitted power $P_\rmt=45$ dBm and the noise of $\sim-84$ dBm, assuming $N_{\rm est}=5$ estimated paths per channel. This setting allows the estimation of a sufficient number of paths with reasonable accuracy to support reliable localization performance.}
		\label{chan_track_err}
	\end{figure*}

	\subsection{F-MOMP for Channel Tracking}
	We consider the communication architecture where a $N_\rmt^\rmx\times N_\rmt^\rmy=16\times 16$ \ac{URA} and four $N_\rmr^\rmx\times N_\rmr^\rmy=12\times 12$ URAs are deployed at the BS and the vehicle, respectively. In every tracking period, the BS transmits $N_\rms=4$ data streams with a length of $Q=36$  drawn from a Hadmard matrix of order $2^6$, with a transmitted power of $P_{\rmt}=45$ dBm. A raised-cosine filter with a roll-off factor of $0.4$ is used as the pulse shaping function. The system operates at the carrier frequency of $f_\rmc=73$ GHz, with a bandwidth of $B_\rmc=1$ GHz. Based on the simulated channel properties and the bandwidth, the number of channel taps is fixed to $N_\rmd=32$. The analog precoders and combiners are constructed based on the historical channel angle estimates, i.e., the beams point toward the directions aligning with the previously estimated \ac{DoA}s and \ac{DoD}s. The vehicle receives $M=40$ measurements to track channel parameters assuming $N_{\rm est}=5$ estimated paths per channel. The resolution for the delay reduced dictionary $\bPsi_1$ is set to $0.25$ ns, and the angular reduced dictionaries $\bPsi_k$ ($k=2,\ldots,5$) are constructed with a resolution of $0.25^\circ$. For all dictionaries, the number of search grids is set to $g_k=8$. Furthermore, the number of iterations for the MOMP algorithm is set to $N_{\rm iter}=4$ to ensure convergence with low computational complexity. The channel tracking results are shown in Fig. \ref{chan_track_err}, where the errors are calculated between the estimated paths and their closest counterparts in the true channel. The delay errors are below $5$ ns for 95\% of the situations, and the 95-th percentile values of the angular errors are $7.1^\circ$, $3.6^\circ$, $2.3^\circ$, and $2.8^\circ$ for the estimated azimuth and elevation \ac{AoA}s, and azimuth and elevation \ac{AoD}s, respectively. The estimation for departure angles has higher accuracy due to the larger antenna array employed at the BS. In addition, paths with higher gain magnitude allow higher angle estimation accuracy, as shown in Fig. \ref{chan_track_err}d, which motivates us to assign weights proportional to the estimated path gains to prioritize more reliable paths during the localization phase to enhance the accuracy. 
	\begin{figure}[!t]
		\centering
		\subfloat[]{%
			\label{voChAT_err_ori}
			\includegraphics[width=0.48\linewidth]{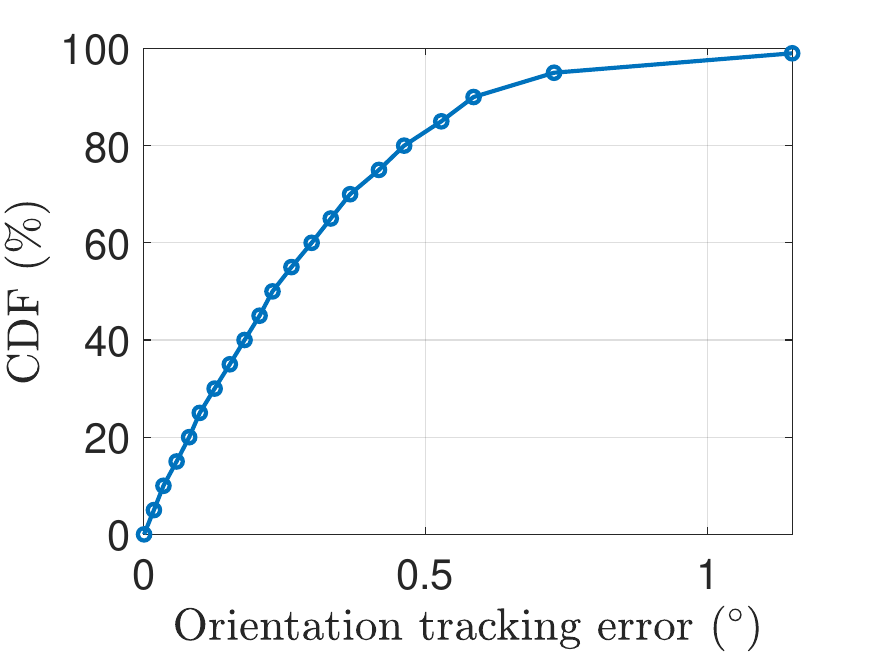}}\hfill
		\subfloat[]{%
			\label{voChAT_err_ss_loc}
			\includegraphics[width=0.48\linewidth]{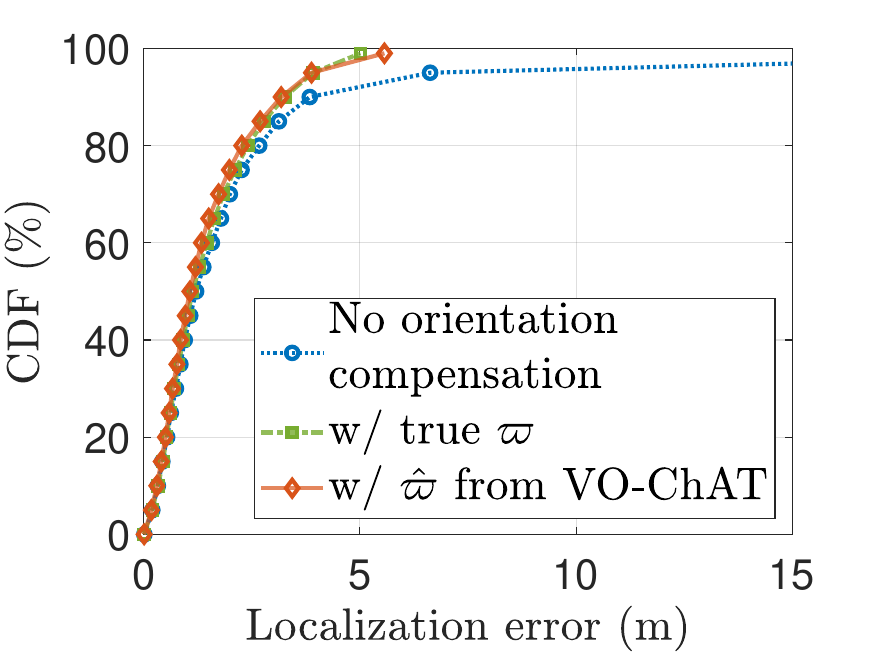}}
		\caption{(a) VO-ChAT for tracking unknown orientation for trajectories in $\cS_{\rm te}$; (b) Single-shot geometric localization results, where using the VO-ChAT predicted orientation estimates $\hat{\varpi}$ achieves comparable performance to that using the true orientation values $\varpi$.}
		\label{voChAT_err}
	\end{figure}
	\subsection{VO-ChAT for Orientation Tracking and Single-Shot Localization after Orientation Compensation}
	We consider the tracking length of $\Gamma=8$, and the input has a dimension of $8\times 5\times 6$. According to the design in Fig. \ref{VPO-ChAT_Arch}, there are six MLP modules in VO-ChAT, denoted as ${\rm MLP}_\sfi$ for $\sfi={\sf 1,...,6}$: four modules within the ChanSTA component, one before and one after the concatenation of historical orientation estimates. Each MLP module consists of dense layers with neuron configurations as $(32, 128)$ for ${\rm MLP}_{\sf 1}$, $(128, 128)$ for ${\rm MLP}_{\sfi}$, $\sfi = {\sf 2, 3, 4}$, $(32, 8, 1)$ for ${\rm MLP}_{\sf 5}$, and $(16, 16)$ for ${\rm MLP}_{\sf 6}$, where each tuple represents the number of neurons per dense layer. The SiLU activation function \cite{elfwing2018sigmoid} is applied after each FC layer to introduce non-linearity into the network. We consider a single head and two heads for the two \ac{MHA} modules, respectively, and the embedding dimensions are set to $K_{\rm dim}=Q_{\rm dim}=32$ for keys and queries, and $V_{\rm dim}=128$ for values, for both the \ac{MHA} modules. Among the 8 trajectories in the database, VO-ChAT is trained on 24 trajectories, denoted as $\cS_{\rm tr}$, and tested on the other 8 trajectories denoted as $\cS_{\rm te}$. The network training employs \ac{MSE} loss with the Adam optimizer for 500 epochs, incorporating early stopping based on validation performance to prevent overfitting. The learning rate is $0.001$ with a decay rate of $0.95$ every $80$ epochs. The orientation tracking performance on the testing set is shown in Fig. \ref{voChAT_err_ori}, where the 50, 80, 95-th percentile errors are $0.23^\circ$, $0.46^\circ$, and $0.73^\circ$, respectively. 
	
	After the orientation compensation to retrieve the angle values in the global coordinate system, we acquire the single-shot geometric localization results employing weighted path contributions, where the weight for each path $\ell$ is computed as $w_{\ell}=|\hat{\alpha}_\ell|-\min\{|\hat{\alpha}_1|,...,|\hat{\alpha}_{|\frak{L}|}|\}+\epsilon_{|\alpha|}$, where $\epsilon_{|\alpha|}=2$ is the positive constant that ensures non-zero weights for all paths. As presented in Fig. \ref{voChAT_err_ss_loc}, the localization errors are below $1.06$ m, $2.26$ m, and $3.88$ m for 50\%, 80\%, and 95\% of the cases, respectively. The results are compared to the situation with no orientation compensation and with perfect orientation knowledge, where using the predicted $\hat{\varpi}$ achieves comparable performance to that with the true $\varpi$, while without orientation compensation the 95-th percentile accuracy is $6.62$ m.
	\begin{table}[!b]
		\centering
		\resizebox{\linewidth}{!}{
			\begin{tabular}{
					>{\columncolor[HTML]{EFEFEF}}c 
					>{\columncolor[HTML]{FFFFFF}}l 
					>{\columncolor[HTML]{FFFFFF}}l 
					>{\columncolor[HTML]{FFFFFF}}l 
					>{\columncolor[HTML]{FFFFFF}}l 
					>{\columncolor[HTML]{FFFFFF}}l |
					>{\columncolor[HTML]{FFFFFF}}l 
					>{\columncolor[HTML]{FFFFFF}}l 
					>{\columncolor[HTML]{FFFFFF}}l }
				\specialrule{.1em}{0em}{0em}
				\cellcolor[HTML]{EFEFEF}                                                & \multicolumn{5}{c|}{\cellcolor[HTML]{FFFFFF}\textbf{Encoder}}                                                        & \multicolumn{3}{c}{\cellcolor[HTML]{FFFFFF}\textbf{Decoder}}             \\ \cline{2-9} 
				\multirow{-2}{*}{\cellcolor[HTML]{EFEFEF}\textbf{Module}}               & ${\rm MLP}_{\sf 1}$ & ${\rm MLP}_{\sf 2}$ & ${\rm MLP}_{\sf 3}$ & ${\rm MLP}_{\sf 4}$ & ${\rm MLP}_{\sf 5}$ & ${\rm MLP}_{\sf 1}$ & ${\rm MLP}_{\sf 2}$ & ${\rm MLP}_{\sf 3}$ \\ \specialrule{.04em}{0em}{0.01em}
				\textbf{\begin{tabular}[c]{@{}c@{}}Layer \\ specification\end{tabular}} & (32, 128)           & (128, 128)          & (128, 128)          & (128, 128)          & (32)                & (8)                 & (32)                & (32)                \\ \specialrule{.1em}{0.01em}{0em}
		\end{tabular}}
		\caption{MLP madule specifications for the encoder and decoder of VP-ChAT.}
		\label{en_de_layers}
	\end{table}
	\begin{figure}[!t]
		\centering
		\subfloat[]{%
			\label{vpChAT_trjExmp}
			\includegraphics[width=0.75\linewidth]{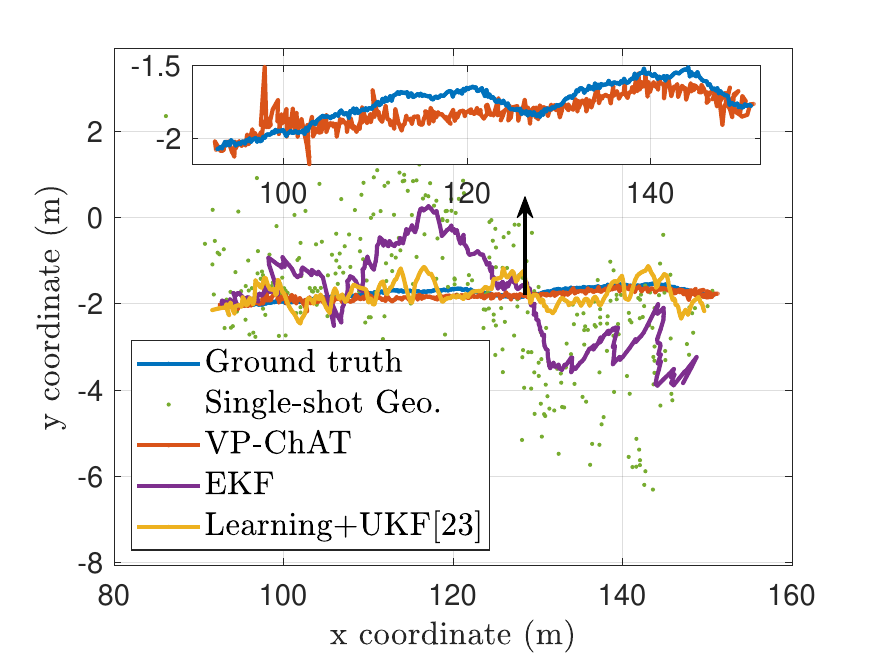}}\\
		\subfloat[]{%
			\label{vpChAT_trjOriExmp}
			\includegraphics[width=0.75\linewidth]{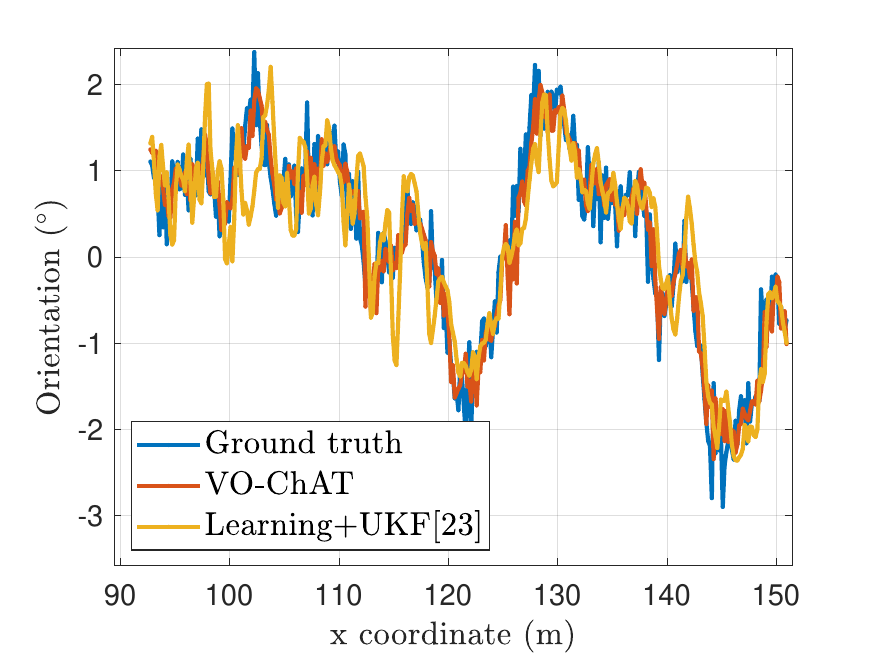}}
		\caption{An example of (a) position tracking performance, and (b) orientation tracking performance based on a trajectory from $\cS_{\rm te}$.}
		\label{trjExmp}
	\end{figure}
	\begin{figure}[!t]
		\centering
		\includegraphics[width=.8\linewidth]{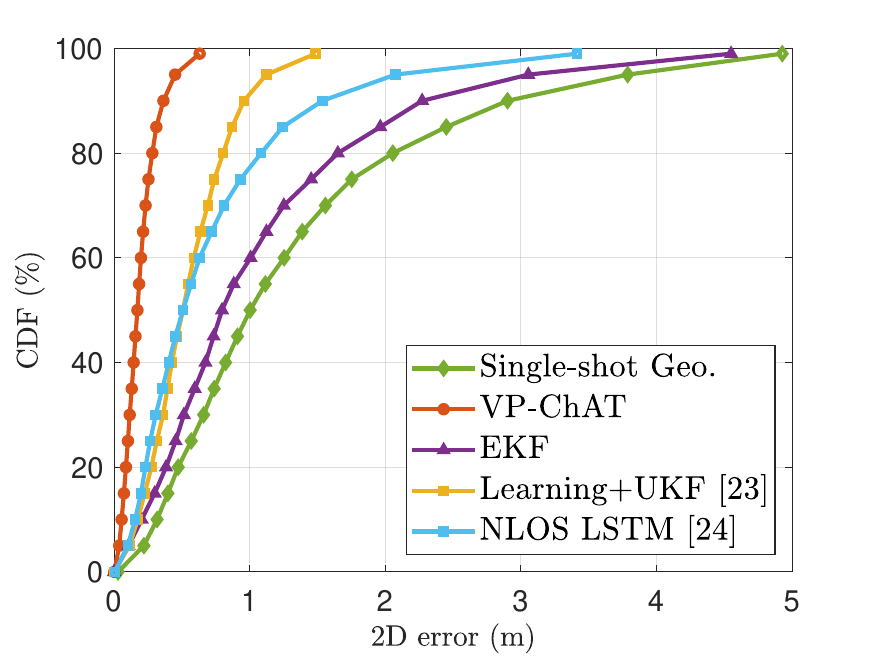}
		\caption{CDF of the position tracking error for VP-ChAT and relevant SOTA methods.  VP-ChAT significantly outperforms prior work, achieving 26 cm accuracy for 80\% of the users.}
		\label{vpChAT_err_locCDF}
	\end{figure}
	\subsection{VP-ChAT for Position Tracking}
	VP-ChAT employs an encoder-decoder architecture where the encoder processes estimated channel information using the same structure as VO-ChAT, i.e., the input channel sequence has a length of $\Gamma=8$ and its dimension is $8\times 5\times 6$, while the decoder analyzes the input position sequence with a length of $\Gamma=8$ to generate position corrections for the current single-shot position estimate. The encoder ahead of the cross-MHA module comprises five MLP modules with layer configurations specified in Table \ref{en_de_layers}, following the same notations defined for VO-ChAT for simplicity. The encoder adopts single-head attention for the first MHA and two-head attention for the second MHA module. Besides, the decoder employs MLP modules with FC layer configurations specified in Table \ref{en_de_layers}, processing position evolution information through single-head self-attention. The following single-head cross-attention between the decoder-generated query and encoder-produced value-key pair is implemented considering the inside MLP with a single layer of $32$ neurons, and the final MLP module consists of a single layer of $8$ neurons. Similar to training VO-ChAT, the Adam optimizer and \ac{MSE} loss are considered for training on $\cS_{\rm tr}$ for 1000 epochs with early stopping. 
	
	An example of tracking performance on a trajectory from $\cS_{\rm te}$ is presented in Fig. \ref{vpChAT_trjExmp}, where the VP-ChAT tracked positions align with the ground truth with deviations of $\leq 0.47$ m, while using \ac{EKF} results in an average tracking error of $0.5$ m and the accuracy of $2.55$ m at the 95th percentile. The position tracking performance based on trajectories in $\cS_{\rm te}$ is demonstrated in Fig. \ref{vpChAT_err_locCDF}. We realize submeter localization across all trajectories, with localization errors below $0.15$ m, $0.27$ m, and $0.43$ m at the 50th, 80th, and 95th percentiles, respectively. For comparison, we implement an \ac{EKF} as the baseline, considering the state vector of $\left[x,y,v_\rmv,\varpi\right]$, and reproduce the algorithm proposed in \cite{Bader2024leverag}, which considers a similar urban driving scenario, addresses clock offset using \ac{RTT}, and identifies higher-order reflections via a learning method trained on 3.6 million data samples, followed by vehicle PO tracking using a UKF. While \cite{Bader2024leverag} assumes idealized channel parameters, our implementation adopts channel estimates obtained through F-MOMP.
	
	\section{Conclusion}\label{conclu}
	We developed a hybrid model/data-driven framework for mmWave communication channel tracking and precise  vehicle PO  tracking in urban environments, adopting realistic system models that account for factors often neglected in prior studies. First, we introduced a low-complexity time domain channel tracking algorithm, F-MOMP, to accurately estimate multipath parameters with delay and angular errors below $0.1$ ns and $2^\circ$ for 80\% of cases, sufficiently supporting vehicle localization. Then, VO-ChAT, employing an attention mechanism to process channel estimate sequences, tracks the vehicle’s orientation with errors below $0.5^\circ$ in 80\% of cases. Thereafter, we formulated a WLS problem  using the selected LOS and first-order channel paths to realize single-shot localization. Finally, VP-ChAT, built upon the Transformer architecture, leverages the channel and position estimation sequence to provide the correction for the single-shot position estimate, achieving the tracking accuracy of $15$ cm and $26$ cm at the 50th and 95th percentiles, respectively. 
	
	The results demonstrate that the hybrid model/data-driven approaches for precise vehicle tracking with mmWave channel estimates in complex urban environments are effective, with deep learning modules integrated when model-based methods exhibit limitations. The network designs are guided by intuitive principles for effective information processing and feature extraction, with the attention mechanism proving its efficacy for accurate results. While large-scale networks used in language models and multimodal information processing consist of billions of network parameters and extensive training data \cite{zhao2024surveylargelanguagemodels}, our streamlined networks efficiently achieve the objectives.

	\bibliographystyle{IEEEtran}
	{
		\scriptsize
		\bibliography{IEEEabrv, refs_abrv}}
	
\end{document}